\documentclass[twocolumn,amsfonts, ammsymb, amsmath, nobibnotes, aps, prd, nofootinbib, superscriptaddress]{revtex4-2}

\usepackage{amsmath}
\usepackage{amsfonts}
\usepackage{amssymb}
\usepackage{booktabs}
\usepackage[T1]{fontenc}
\usepackage{graphicx}
\usepackage[colorlinks=true, linkcolor=blue, urlcolor=blue, citecolor=blue]{hyperref}
\usepackage{siunitx}
\usepackage{xcolor}
\usepackage{orcidlink}

\usepackage{graphicx}	
\usepackage{amsmath}	
\usepackage{amssymb}	
\usepackage{siunitx}
\usepackage{xcolor}
\usepackage{subcaption}

\DeclareRobustCommand{\VAN}[3]{#2}
\let\VANthebibliography\thebibliography
\def\thebibliography{\DeclareRobustCommand{\VAN}[3]{##3}\VANthebibliography}

\newcommand{\uib}{IAC3, Universitat de les Illes Balears, Cra. de Valldemossa km 7.5, 07122 Palma, Spain}

\begin{document}

\title{Prospects for Simultaneously Detecting Long and Short Duration Gravitational Waves from Pulsar Glitches with Current and Future Detectors}

\author{M.~Ball\orcidlink{0000-0001-5565-8027}}
\email{matthew.ball@ligo.org}
\affiliation{\uib}

\author{D.~Keitel\orcidlink{0000-0002-2824-626X}}
\affiliation{\uib}

\begin{abstract}
The internal mechanisms that drive the sudden change in rotation frequency in pulsar glitches could emit gravitational waves (GWs) at multiple time scales. The potential for GW analyses to detect both long and short forms of emission (``tCW-like'' and ``burst-like'') improves as the detectors increase in sensitivity. We aim to examine how future detectors fare in making this joint measurement. By examining past glitches from the ATNF catalog, we quantify the double-detectability prospects of current- and next-generation GW detector networks. We further explore what astrophysical constraints could be placed following a double non-detection using extrapolations from the 2024 Vela pulsar glitch as a stand-in example. We find that a network with at least one next-generation detector greatly increases the probability of double-detection for many known glitching pulsars and improves the detection threshold in glitch energy by nearly two orders of magnitude over existing LIGO detectors. Additionally, double non-detections with networks with at least one next-generation detector could place dramatically improved limits on properties of the pulsar.
\end{abstract}

\maketitle

\section{Introduction}
\label{sec:introduction}
Pulsars are a subclass of rotating neutron stars (NSs) that emit periodic pulses of electromagnetic radiation \citep{lorimer_handbook_2004}. The timing of these pulses is tied to the rotation rate of the pulsar, and observations have shown that, generally, pulsars are slowing down over time. However, some pulsars demonstrate sudden changes in their rotation rate known as ``glitches'' \citep{link_pulsar_1992,cheng_probing_2000,fuentes_glitch_2017}. A glitch consists of a sudden change in rotation rate (timescale of a few seconds) usually followed by a slower relaxation back to the original rotation rate (timescale of days to months) \citep{antonopoulou_pulsar_2022}. These glitches have not yet been tied to a specific mechanism, but models suggest that they may be due to some internal exchange of angular momentum (see eg.~\cite{haskell_models_2015} and~\cite{zhou_pulsar_2022} for reviews).

NSs are believed to contain exotic matter potentially exceeding nuclear densities, but their specific internal structure is poorly understood. Thus, pulsar glitches represent an interesting opportunity to probe the interior of these objects through the potential emission of Gravitational Waves (GWs) \citep{haskell_probing_2017}. Glitches are considered potential sources of GWs for ground-based laser-interferometric detectors such as LIGO~\citep{KAGRA:2013rdx,LIGOScientific:2014pky}, Virgo~\citep{VIRGO:2014yos}, KAGRA~\citep{10.1093/ptep/ptaa125}, Cosmic Explorer (CE)~\citep{evans_horizon_2021,srivastava_science-driven_2022}, and the Einstein Telescope (ET)~\citep{regimbau_mock_2012, regimbau_mock_2026, meacher_einstein_telescope_2014, branchesi_science_2023, ET_coba}. Searches for GWs associated with pulsar glitches have already been conducted as far back as the initial LIGO era~\citep{abadie_search_2011}. Analyses have targeted both the extended relaxation time as a source of longer transient Continuous Waves (tCWs) \citep{abadie_search_2011,prix_search_2011,keitel_first_2019,lopez_prospects_2022,lopez_search_2024,O4_vela_glitch_search} and the initial jump as a source of short-duration GWs (referred to as ``bursts'') \citep{abbott_all-sky_2017,abbott_all-sky_2019,abbott_all-sky_2021,abac_all-sky_2025}. Of note is that the most recent analysis of the 2024 Vela pulsar glitch included searches for both tCWs and bursts \citep{O4_vela_glitch_search}.

Due to the potential for multiple mechanisms of GW emission, in this paper, we consider both long duration tCWs and short duration bursts in tandem.~\cite{ball_methods_2025} showed that a combined analysis of the two GW channels can lead to improved constraints on NS source parameters. The aim of this work is to estimate how this framework extends to planned future GW detectors and estimate the sensitivity of future analyses to measure both long and short duration emission simultaneously (dual detection). The dual detection of both tCWs and bursts was the ideal outcome of the analyses in~\cite{O4_vela_glitch_search}, but they were unsuccessful. While~\cite{ball_methods_2025} did consider how sensitive future detectors would be to glitches like the 2016 Vela pulsar glitch, we aim to expand this to a larger pulsar glitch population to make broader statements about potential targets of future searches. Additionally, improvements in radio observations with the Square Kilometer Array are expected to nearly quadruple the number of known pulsars by the time that third-generation GW detectors come online~\citep{keane_square_2025}, so there may be even more potential targets for next-generation detectors.

This paper is organized as follows: in Section~\ref{sec:GWs}, we describe the models for GW emission in each emission channel. Then, in Section~\ref{sec:searches}, we review the search methods used to search for each type of signal. We describe how we estimate the sensitivity of different detector networks to these types of analyses in Section~\ref{sec:sampling} and review the catalog of pulsar glitches used for this analysis in Section~\ref{sec:Pulsar_catalog}. In Section~\ref{sec:methods}, we detail our methodology for this analysis. Finally, in Section~\ref{sec:Results}, we check detectability prospects for the full set of pulsar glitches considered, then specialize to the case of the Vela pulsar and study the improvement on astrophysical constraints that can be placed based on double non-detections.

\section{Gravitational Wave Emission Models}
\label{sec:GWs}
Our analysis relies on a coherent description of both the long duration tCW emission and the short duration burst emission. Multiple models exist for both channels, but for simplicity, we choose the slowly damped, transient mountain model for tCW emission and the short-duration f-mode for burst emission. Our prescription is similar to the one used in~\cite{ball_methods_2025}.

\subsection{Continuous Gravitational Waves}
\label{subsec:CW_model}

While other models for tCW emission exist \citep[e.g.][]{van_eysden_gravitational_2008,bennett_continuous-wave_2010,melatos_persistent_2015,singh_gravitational_2017} the most intuitive tCW mechanism is the transient mountain model \citep{yim_transient_2020} where some unknown internal mechanism induces an ellipticity in the pulsar which is damped out through GW emission over the long recovery.

In this model, the emission is characterized by a slowly damped sinusoid parameterized by a damping time $\tau_{\mathrm{CW}}$, a frequency $\nu_{\mathrm{CW}}$, and an amplitude $h_{0,\mathrm{CW}}$. For the simplest form of tCW emission, the frequency is twice the pulsar rotation speed $\nu_{\mathrm{CW}}=2\nu_{\mathrm{s}}$, and the damping time $\tau_{\mathrm{CW}}$ is expected to be related to the glitch recovery time by $\tau_{\mathrm{CW}} = 2\tau_{\mathrm{glitch}}$ \citep{yim_transient_2020}.

Given some energy budget $E_{\mathrm{CW}}$ and integrating the GW luminosity as a function of time for a slowly damped sinusoid waveform, the amplitude can be written as 
\begin{equation}
    h_{0,\mathrm{CW}} = \frac{1}{2 \pi d \nu_{s}} \sqrt{\frac{5G E_{\mathrm{CW}}}{c^3 \tau_{\mathrm{CW}}}},
    \label{eq:CW_amplitude}
\end{equation}
where $d$ is the distance to the source and $E_{\mathrm{CW}}$ is the energy carried by the tCW.

The characteristic glitch energy $E_{\mathrm{glitch}}$ is given by 
\begin{equation}
    E_{\mathrm{Glitch}} = 4 \pi^2 \mathcal{I}_{zz} \nu_{s} \Delta \nu_{s},
    \label{eq:glitch_energy}
\end{equation}
where $\Delta \nu_{s}$ is the change in rotation rate associated with the glitch and $\mathcal{I}_{zz}$ is the moment of inertia.

The moment of inertia can be expressed in terms of the mass and radius of the neutron star via the equation-of-state independent fitting function from~\cite{yang_exploring_2022} :
\begin{equation}
    \log\left(\frac{I_{zz}}{M^3}\right) = -1.261\log\left(\frac{M}{R}\right)+0.277.
    \label{eq:inertial_fitting_function}
\end{equation}
This is an updated relation from the one from~\cite{yagi_approximate_2017} used in~\cite{ball_methods_2025} and~\cite{O4_vela_glitch_search}.

The energy available to be emitted as a tCW is not necessarily equal to the full glitch energy. Since only some portion of the total spin change $\Delta \nu_{s}$ is recovered after the glitch,~\cite{yim_transient_2020} introduced a ``healing parameter'' $Q$ that quantifies the fraction of the glitch spin change that is recovered. They argued that this fraction could be the fraction of energy emitted as a tCW. For consistency with~\cite{ball_methods_2025}, we denote such a fraction as $\digamma_{\!\mathrm{CW}}$ and write the tCW energy as
\begin{equation}
    E_{\mathrm{CW}} = \digamma_{\!\mathrm{CW}} E_{\mathrm{Glitch}}.
    \label{eq:CW_energy}
\end{equation}
We will later fix $\digamma_{\!\mathrm{CW}} = Q$.

\subsection{Gravitational Wave Bursts}
\label{subsec:Burst_model}

The most generally accepted form of a gravitational wave burst from a pulsar glitch is the f-mode. It is described by a rapidly damped sinusoid parameterized by an amplitude $h_{0,f}$, a frequency $\nu_{f}$, and a damping time $\tau_{f}$~\citep{echeverria_gravitational-wave_1989,finn_detection_1992}:

\begin{equation}
h(t) = 
    \begin{cases}
        0 & \text{for } t < 0\\
        h_{0,f} e^{-t/\tau_{f}} \sin{\left(2\pi \nu_{f}t\right)} & \text{for } t \geq 0.
    \end{cases}
    \label{eq:fmode_damped_sinusoid}
\end{equation}

As in~\cite{ball_methods_2025}, we use equation-of-state independent fitting functions from~\cite{pradhan_general_2022} to relate the frequency (Equation~\ref{eq:fmode_frequency}) and damping time (Equation~\ref{eq:fmode_damping}) to the neutron star mass and radius (here $\overline{M} = M/1.4 M_{\odot}$ and $\overline{R} = R/10 \: \mathrm{km}$) 

\begin{equation}
    \nu_{f,0}(\mathrm{kHz})\approx0.535+1.646\left(\frac{\overline{M}}{\overline{R}^3}\right)^{1/2}
    \label{eq:fmode_frequency}
\end{equation}

\begin{equation}
\frac{1}{\tau_f(s)}\approx \frac{\overline{M}^3}{\overline{R}^4}\left[21.19-13.41\left(\frac{\overline{M}}{\overline{R}}\right)\right]
\label{eq:fmode_damping}
\end{equation}

Since pulsars are rotating with a known frequency, we can include a correction to the f-mode frequency to account for this \citep{kruger_fast_2020}. This is shown in Equation~\ref{eq:fmode_rotation} below, where $\nu_{f}$ is the observable f-mode frequency, $\nu_{f,0}$ is the f-mode frequency for an equivalent non-rotating NS (Equation~\ref{eq:fmode_frequency}), and $\nu_s$ is the rotation frequency of the neutron star, all in units of kHz.

\begin{equation}
    \nu_f = \nu_{f,0}\left(1+0.220\left(\frac{\nu_s}{\nu_{f,0}}\right)-0.0170\left(\frac{\nu_s}{\nu_{f,0}}\right)^2\right)
    \label{eq:fmode_rotation}
\end{equation}

The amplitude of the f-mode is obtained similarly to the transient mountain case. By integrating the GW luminosity as a function of time for a damped sinusoid, we can write the f-mode amplitude as 
\begin{equation}
    h_{0,f} = \frac{1}{\pi d \nu_{f}} \left(\frac{5G}{c^3}\frac{E_{GW,f}}{\tau_{f}}\right)^{1/2},
    \label{eq:h0_fmode}
\end{equation}
where $E_{GW,f}$ is the energy in the f-mode channel.

We relate the f-mode energy to the glitch energy as in~\cite{ball_methods_2025} with some fraction $\digamma_{\!\mathrm{fmode}}$:

\begin{equation}
    E_{\mathrm{fmode}} = \digamma_{\!\mathrm{fmode}} E_{\mathrm{Glitch}}.
    \label{eq:fmode_energy}
\end{equation}

\section{Gravitational Wave Searches}
\label{sec:searches}
The two types of GW signals considered in this analysis require different approaches to identify in the data. Longer duration tCWs require a more sensitive matched-filter approach to dig through the noise, and shorter duration bursts require a minimally modeled approach due to the relative ambiguity in the signal morphology.

\subsection{tCW Searches}
\label{subsec:CW_search}

Targeted searches for tCWs consider narrowband GWs at $1\times$ and $2\times$ the pulsar rotation frequency. They assume that the signal frequency is relatively constant and integrate the detector data over the timescale of days to months. Here, we consider the template-based, matched-filter transient $\mathcal{F}$-statistic algorithm (implemented in the \textsc{Lalsuite} code base \citep{lalsuite_2018}).

This analysis identifies tCW candidates via a maximum likelihood method and estimates the dimensionless GW amplitude $h_{0,\mathrm{CW}}$ \citep{jaranowski_data_1998,prix_search_2011}. If no signal is detected, injections are performed to estimate the maximum sensitivity of the search. Signals of varying amplitude and duration are injected and recovered to estimate what values of $h_{0,\mathrm{CW}}$ the search is sensitive to. The amplitude for which $95\%$ of signals are estimated to be recoverable is typically reported as the upper limit of sensitivity for the search.

\subsection{Burst Searches}
\label{subsec:Burst_search}

Searches for short-duration bursts are typically minimally modeled, relying on coherent statistics instead of templates to identify candidates. These searches generate time-frequency maps of some coherence statistic using data from multiple detectors. These maps consist of time-frequency ``pixels'' assigned a coherent energy, and adjacent pixels with the highest energy are grouped together into GW candidates. Burst searches typically target frequencies ranging from $100\,\mathrm{Hz}$ to $4000\,\mathrm{Hz}$. The two most commonly used algorithms for this type of analysis are \textsc{Coherent WaveBurst} \citep{drago_coherent_2021,lopez_prospects_2022} and \textsc{X-Pipeline} \citep{sutton_x-pipeline_2010,was_performance_2012}.

Candidate events and upper limits are reported with the integrated root-sum-squared amplitude $h_{\mathrm{rss}}$:

\begin{equation}
    h_{\mathrm{rss}}^2 = \int_{-\infty}^{\infty}\left|\tilde{h}_{+}(f)\right|^2 + \left|\tilde{h}_{\times}(f)\right|^2 \mathrm{df}.
    \label{eq:hrss}
\end{equation}

In a non-detection, phenomenological waveforms are injected into the data at varying amplitudes, frequencies, and durations to estimate the sensitivity of the search, reported as the $h_{\mathrm{rss}}$ value at which at least $90\%$ of injections are recovered, referred to as the ``detection efficiency''. In this analysis, we only consider ``ringdown'' waveforms (damped sinusoids with a smooth ramp-up at the beginning) to represent f-modes as in~\cite{ball_methods_2025}.

\section{Sensitivity Limits}
\label{sec:sampling}
In order to make statements on the detectability of GWs from pulsar glitches, we to be able to estimate the sensitivity of future detectors based on current detector sensitivity. The two different GW search techniques require different approaches to estimating future sensitivity. tCW searches integrate long periods of time, so we need to properly consider averages of detector data over noise artifacts. Short-duration burst searches are highly sky location dependent, and we must take care to properly fold in the impact of the detector antenna patterns, target sky location, and the search time.

We will later consider 5 total detector networks: the two current LIGO detectors (LHO \& LLO) at O4 sensitivity~\citep{capote_advanced_2025}, LHO \& LLO at O5 sensitivity~\citep{Aplus_GWINC_2018,Aplus_GWINC}, LHO \& LLO at O5 sensitivity with CE (assumed as a single $40\,\rm{km}$ L-shaped detectors)~\citep{evans_horizon_2021,srivastava_science-driven_2022}, LHO \& LLO at O5 sensitivity with ET (treating ET as a combination of 3 detectors in the triangular configuration~\citep{regimbau_mock_2012, regimbau_mock_2026, meacher_einstein_telescope_2014, branchesi_science_2023, ET_coba} with parameters defined in~\cite{lalsuite_2018}), and a 3rd-generation network with CE and ET. For O4 sensitivity, we only consider the LHO and LLO detectors since Virgo was not used in the 2024 Vela glitch analysis~\cite{O4_vela_glitch_search}. For O5, we also restrict our analysis to the LIGO detectors for consistency even though the addition of Virgo and KAGRA to the network will further improve search sensitivity~\citep{KAGRA:2013rdx}. We also note that the ET geometry is still under consideration. If it is changed from a triangular geometry to two L-shape interferometers, the sensitivity is expected to be more in line with that of CE~\citep{branchesi_science_2023,bruno_einstein_2026}. We use the fiducial locations of CE and ET as implemented in \textsc{Lalsuite}~\citep{lalsuite_2018} to estimate antenna patterns. This places CE aligned with the LHO detector, and ET at the location of the Virgo detector.

\subsection{tCW Sensitivity}
\label{subsec:CW_ULs}

To estimate the sensitivity of tCW searches for current and future detectors, we need to take into account the source sky location and orientation, detector antenna patterns, detector locations, and observing time. We follow the formalism in~\cite{prix_2011} to define the optimal signal-to-noise ratio $\rho(0)$ in terms of a (non-transient) CW amplitude $h_0$ (Equation 77 in~\cite{prix_2011}):

\begin{equation}
    \rho^2(0) = h_0^2 \left(\alpha_1 A + \alpha_2 B + 2 \alpha_3 C + 2 \alpha_4 E \right)\mathcal{S}^{-1}T_{\mathrm{data}}.
    \label{eq:CW_optimal_snr}
\end{equation}

Here, $\alpha_1$, $\alpha_2$, $\alpha_3$, and $\alpha_4$ are functions of the source inclination $\iota$ and phase $\psi$ defined in Equations 78-81 in~\cite{prix_2011}. $A$, $B$, $C$, and $E$ are elements of the antenna pattern matrix defined in Equation 66 in~\cite{prix_2011}. $\mathcal{S}$ is the harmonic mean of the noise power spectral densities (PSDs) of the detectors in the network:
\begin{equation}
    \mathcal{S}^{-1} = \frac{1}{N_{\mathrm{det}}}\sum_{X} S_X^{-1}.
    \label{eq:Sinv_def}
\end{equation}
This definition is similar to Equation 60 in~\cite{prix_2011}, but we are directly using run-averaged (or predicted) PSDs $\sqrt{S_X}$ for each detector $X$ instead of discrete SFTs.

Furthermore, since our implementation does not use SFTs, our definition of $T_{\mathrm{data}}$ is a simplification of Equation 64 in~\cite{prix_2011} to $T_{\mathrm{data}}\equiv N_{\mathrm{det}}\frac{T_{\mathrm{obs}}}{T_{\mathrm{SFT}}} \mu$, 

where $N_{\mathrm{det}}$ is the number of detectors in the network, $T_{\mathrm{obs}}$ is the effective observation time (used interchangeably with the tCW signal duration in our case), $T_{\mathrm{SFT}}$ is the duration of SFTs for a simulated search, and $\mu$ is the duty factor for coincident observation time. $T_{\mathrm{data}}$ is the effective total coherent observing time by the detector network in Equation \ref{eq:CW_optimal_snr}. We let $T_{\mathrm{SFT}}=1800\,\mathrm{s}$ and $\mu=0.66$ to match~\cite{moragues_prospects_2023}.

In practice, these terms are computed with various functions from \textsc{lalsuite}~\citep{lalsuite_2018,swiglal_2020}. For a given network, we take the detector sensitivity curves and compute detector noise weights with Equation 58 in~\cite{prix_2011} for time stamps corresponding to tiling the signal duration with SFTs. We also use \textsc{lalpulsar}'s \textsc{GetMultiDetectorStates} function (as implemented in~\cite{pyfstat}) to compute a network detector state time series given parameters of the detectors in the network and the signal duration. The weights and detector states are used to compute the parameters $A$, $B$, $C$, and $E$ with \textsc{lalpulsar}'s \textsc{ComputeMultiAMCoeffs} function.\footnote{The normalization of the coefficients output by \textsc{ComputeMultiAMCoeffs} differ from those in Equation~\ref{eq:CW_optimal_snr} by a factor of $N_{\mathrm{SFT}}$.}

To estimate sensitivity, we need to set a threshold on $\rho$ in Equation~\ref{eq:CW_optimal_snr}. In practice, the detectability threshold of a search depends on many aspects of the search setup and data quality. As a simple but realistic estimate, we reuse the detection statistic threshold of $2\mathcal{F}=50$ derived in~\cite{moragues_prospects_2023} for mock searches for the Vela or Crab pulsar over realistic parameter-space ranges, and translate it to a SNR threshold via setting $\rho_{\mathrm{thresh}} = \left\langle2\mathcal{F}\right\rangle-4$ (based on the $\chi^2$ distribution of the $\mathrm{F}$-statistic~\cite{prix_2011}). With all of these terms, we rearrange Equation~\ref{eq:CW_optimal_snr} to solve for $h_0$ to use as our threshold for the frequency of the particular pulsar being analyzed.

This process can be computationally expensive when performed many times. To speed things up, we split our process into two scenarios: when $T_{\mathrm{obs}}=\tau_{\mathrm{CW}}\geq1\,\mathrm{day}$ and $T_{\mathrm{obs}}=\tau_{\mathrm{CW}}<1\,\mathrm{day}$. When $T_{\mathrm{obs}}<1\,\mathrm{day}$, we perform the full process depicted above, but when $T_{\mathrm{obs}}\geq1\,\mathrm{day}$, we can take advantage of the fact that the sky location is being averaged over more than one day and re-use the noise weights and detector states for a fixed reference time of 5 days, and re-weight the threshold $h_0$ by a factor of $\sqrt{\tau_{\mathrm{ref}}/\tau_{\mathrm{CW}}}$. Here, we use $\tau_{\mathrm{ref}}=5\,\mathrm{days}$. Since all of this infrastructure was computed for a constant-amplitude signal, and our model is a slowly decaying one, we include an additional factor of $\sqrt{2}$ to account for this difference~\cite{moragues_prospects_2023}.

When making projections of existing $h_{0,\mathrm{CW}}$ values to future detectors, the process is mostly the same. We compute the same detector-dependent parameters for the new network. We assume that a search in the future network will place limits at the same $\rho_{\mathrm{thresh}}$ as the old network, so we can set the $\rho_{\mathrm{thresh}}$ in Equation~\ref{eq:CW_optimal_snr} for the old network equal to the $\rho_{\mathrm{thresh}}$ in the new network and solve for the new amplitude $h_{0,a}$ as a function of the old amplitude $h_{0,b}$:

\begin{widetext}
\begin{equation}
    h_{0,a} = h_{0,b} \sqrt{\frac{\left(\alpha_{1,b} A_{b} + \alpha_{2,b} B_{b} + 2 \alpha_{3,b} C_{b} + 2 \alpha_{4,b} E_{b} \right)\mathcal{S}^{-1}_{b}T_{\mathrm{data},b}}{\left(\alpha_{1,a} A_{a} + \alpha_{2,a} B_{a} + 2 \alpha_{3,a} C_{a} + 2 \alpha_{4,a} E_{a} \right)\mathcal{S}^{-1}_{a}T_{\mathrm{data},a}}}.
    \label{eq:CW_amplitude_projection}
\end{equation}
\end{widetext}

This $h_{0,a}$ value is computed only for the specific frequency of the pulsar being analyzed. We also note that in the case of two networks of only one detector with identical orientations and identical source location, the square root term reduces to the square root of the ratio of the PSDs at the expected GW frequency.

\subsection{Burst Upper Limits}
\label{subsec:Burst_ULs}

For the burst upper limits, we begin with the \textsc{X-Pipeline} upper limits, corresponding to the $90\%$ detection probability\footnote{For clarity, we note that in the Burst community, usually a distinction is drawn between detection efficiency and detection probability. The former considers all detectable signals within the available data and analysis window while the latter corrects this with the duty factor to account for gaps in the data. For these search results, these are directly equivalent since the analysis window was gap-free. This is not necessarily true for other short-duration searches that may have had data gaps in their analysis window.}, reported in the targeted search for GWs from the April 2024 Vela Glitch in~\cite{O4_vela_glitch_search,vela_glitch_zenodo} as this represents the most recent targeted search for GW bursts at this time. We specifically use the ringdown waveform upper limits as described in Section~\ref{subsec:Burst_search} (and Section 4.5 of~\cite{O4_vela_glitch_search}). These results were only for LHO and LLO, as Virgo had very poor sensitivity to the sky location of the Vela pulsar at the time of the glitch~\cite{O4_vela_glitch_search}.

However, these results are for a single sky location at O4 sensitivity at a given time. We can rescale these $h_{\mathrm{rss}}$ limits to other detectors and sky locations and glitch times using the relations described in Section 3 of~\cite{sutton_rule_2013}. Sky location and signal start time are partially degenerate, so this is technically overdefined, but we take this approach for completeness.

For a short, narrowband signal, the relationship between the $h_{\mathrm{rss}}$ and the matched-filter signal-to-noise ratio (SNR) $\rho$ is given by.

\begin{equation}
    \rho^2 = \Theta^2 \frac{h_{\mathrm{rss}}^2}{S_n(f_0)}.
    \label{eq:hrss_to_snr}
\end{equation}

Here, $S_n(f_0)$ is the detector PSD evaluated at the signal frequency $f_0$, and $\Theta$ is a source-dependent parameter defined in Equation~\ref{eq:thetasquare} below. $\theta$, $\phi$, $\psi$, and $\iota$ are angular parameters for right ascension, declination, polarization, and inclination, respectively. $F_{+}$ and $F_{\times}$ are the detector antenna response functions for plus and cross polarizations, respectively. For brevity, we do not explicitly list the dependence on $\theta$, $\phi$, $\psi$, $\iota$ and $t$ in $\Theta$. Our f-mode source model is elliptically polarized.

\begin{widetext}
\begin{equation}
    \Theta^2 \equiv
        \begin{cases} 
            F_{+}^2(\theta,\phi,\psi,t) + F_{\times}^2(\theta,\phi,\psi,t) & \mathrm{Unpolarized} \\
            F_{+}^2(\theta,\phi,\psi,t) \left(\frac{1+\cos^2(\iota)}{2}\right)^2 + F_{\times}^2(\theta,\phi,\psi,t)\cos^2(\iota) & \mathrm{Ellipitical} \\
            F_{+}^2(\theta,\phi,\psi,t) \, 2 \sin^4(\iota) &  \mathrm{Linear}
        \end{cases}
    \label{eq:thetasquare}
\end{equation}
\end{widetext}

For a search performed with some network of detectors, the network SNR is given by the quadrature sum of the detector SNRs: $\rho_{\mathrm{net}}^2 = \sum_{i=1}^{N_{\mathrm{det}}^{\mathrm{net}}} \rho_i^2$. As the quantity $h_{\mathrm{rss}}$ is detector-independent, we can write the network SNR as 
\begin{equation}
    \rho_{\mathrm{net}}^2 = h_{\mathrm{rss}}^2 \sum_{i=1}^{N_{\mathrm{det}}^{\mathrm{net}}}\frac{\Theta_i^2}{S_{n,i}(f_0)},
    \label{eq:network_SNR_to_hrss}
\end{equation}
where the summation index $i$ indicates different detectors in the network, and $N_{\mathrm{det}}^{\mathrm{net}}$ is the number of detectors in the network.

We make the assumption that a pipeline will set upper limits at the same network SNR for any given detector network. This assumes that the new network will be impacted by non-Gaussian noise in the same way as the original, even if it has a different number of detectors. To estimate the $h_{\mathrm{rss,B}}$ for a network $B$ given the $h_{\mathrm{rss,A}}$ for network $A$, we can simply equate the network SNRs and solve for the desired $h_{\mathrm{rss}}$:

\begin{equation}
    h_{\mathrm{rss,B}} = h_{\mathrm{rss,A}}\sqrt{\frac{\sum_{i=1}^{N_{\mathrm{det}}^{\mathrm{A}}}\frac{\Theta_{i}^2}{S_{n,i}(f_0)}}{\sum_{j=1}^{N_{\mathrm{det}}^{\mathrm{B}}}\frac{\Theta_{j}^2}{S_{n,j}(f_0)}}}.
    \label{eq:scale_hrss}
\end{equation}

We note that in the case of two networks of one detector and identical sky locations and times, this relationship simplifies to the ratio of the individual detector sensitivities. With this prescription, we can rescale the 2024 Vela glitch upper limits to any other sky location and time at O4 sensitivity or for any future detector network.

\section{Glitching Pulsar Population}
\label{sec:Pulsar_catalog}
While an ideal analysis would attempt to accurately model the galactic population of pulsar glitches, we choose to use measured glitches from known catalogs to act as a representative sample of the broader population as was done in~\cite{moragues_prospects_2023}. Due to the nature of our model, we need to restrict ourselves to pulsar glitches with a measured healing parameter Q as defined by~\cite{yim_transient_2020}. The ATNF glitch catalog includes this crucial parameter for many of the known glitches \citep{manchester_australia_2005,ATNF_catalog}. This restriction leaves us with 114 catalog glitches.

We also include three additional glitches not in the ATNF catalog that have measured healing parameters: the 2016 glitch of J1740-3015 \citep{lower_utmost_2020}, the 2021 Vela pulsar glitch \citep{zubieta_first_2023}, and the 2024 Vela pulsar glitch \citep{atel16608,zubieta_study_2025} with the inferred healing parameter from~\cite{O4_vela_glitch_search}. This brings the total number of pulsar glitches considered in this analysis to 117 glitches from 57 unique pulsars.

We augment this list with additional information on pulsar orientation (inclination angle $\iota$ and polarization angle $\psi$) with estimates from measurements of wind tori from~\cite{ng_fitting_2008}. Not all the pulsars in our sample have this information available. Those that do not, we will treat as uninformed for these parameters. These orientation parameters are used in both tCW and burst calculations.

\section{Sampling Implementation}
\label{sec:methods}
The general workflow for this analysis is as follows. For an individual pulsar glitch, we perform Monte Carlo simulations of the source parameters. These parameters are used to compute search-specific amplitudes to compare to search limits. We count what fraction of the samples fall above a ``detection threshold'' and report that as a ``detection probability.''

We draw samples from uniform distributions on the mass and radius of the neutron star, set at $[0.9M_{\odot},1.6M_{\odot}]$ and $[8.0 \mathrm{km},16.0\mathrm{km}]$, respectively. Source distributions for the rotation rate $\nu_s$, the glitch-induced change in rotation rate $\Delta\nu_s$, the glitch recovery time $\tau_{\mathrm{CW}}$, the right ascension $\theta$, and the declination $\phi$ are Gaussian distributions with mean and $1\sigma$ ranges taken from the literature (see Section~\ref{sec:Pulsar_catalog}). Unless specified otherwise, the distances $d$ to each pulsar are fixed to literature values. We uniformly draw samples from a 24-hour window on glitch time to include the effects of Earth's rotation on sky location. For glitches from pulsars where the measurements are available, we draw inclination $\iota$ and polarization $\psi$ from Gaussian distributions informed by the literature \citep{ng_fitting_2008}, otherwise, the uninformed distributions are $\sin(\iota)\in[-1,1]$ and $\psi\in[0,2\pi]$. Following the assumption in Section~\ref{subsec:CW_model}, we set $\digamma_{\!\mathrm{CW}}=Q$, so the distribution for $\digamma_{\!\mathrm{CW}}$ is a Gaussian distribution based on the literature value of the healing parameter $Q$.

We also constrain our samples to those which only produce f-modes with frequencies between $800\,\mathrm{Hz}$ and $3200\,\mathrm{Hz}$ and damping times greater than $0\,\mathrm{s}$ and less than $10\,\mathrm{s}$~\citep{andersson_towards_1998}. This primarily affects the mass-radius parameter space. These are further constrained to values that obey causality \citep{buchdahl_general_1959}.

Values for $\digamma_{\!\mathrm{fmode}}$ is less astrophysically motivated than $\digamma_{\!\mathrm{CW}}$, so we do not assume any one specific value. We consider 2 different assumptions for $\digamma_{\!\mathrm{fmode}}$ relative to the value of $\digamma_{\!\mathrm{CW}}$:
\begin{itemize}
    \item $\digamma_{\!\mathrm{fmode}}=1-Q$ such that the f-mode and tCW energy always add up to the characteristic glitch energy
    \item $\digamma_{\!\mathrm{fmode}} = \digamma_{\!\mathrm{CW}} = Q$ such that both the f-mode and tCW energy are always equal
\end{itemize}

We choose these values for our parameter space since $Q$ tends to lower values, this shows the limits of large and small values of $\digamma_{\!\mathrm{fmode}}$

From the set of samples drawn from these distributions, we compute the tCW $h_{0,\mathrm{CW}}$ (Equation~\ref{eq:CW_amplitude}) and the f-mode $h_{\mathrm{rss}}$ (Equation~\ref{eq:hrss}). These are compared to the detection statistics described in Sections~\ref{sec:searches} and~\ref{sec:sampling}.

We treat the upper limit thresholds placed by searches as the cutoff for detection in our analysis. For generic CW searches, the standard threshold is the $h_{0,\mathrm{CW}}$ amplitude at which $95\%$ of injections are recovered. For burst searches, the standard threshold is the $h_{\mathrm{rss}}$ at which $90\%$ of injections are recovered. These thresholds can be interpreted as the amplitude at which a real signal has a $95\%$ or $90\%$ probability of being recovered by the CW or burst search, respectively. If the amplitude of a sample drawn from our source distributions falls above the appropriate threshold, we consider that sample as ``detectable.'' This is done for all samples drawn for the glitch, and the total fraction of samples that are detectable by the CW search, the burst search, and both searches simultaneously (both $h_{0,\mathrm{CW}}$ and $h_{\mathrm{rss}}$ fall above the respective thresholds) are recorded as ``dual-detectable.'' This process is repeated for each pulsar.

\section{Observational Prospects}
\label{sec:Results}
\subsection{Population Prospects}
\label{subsec:pop_prospects}

\begin{figure}
    \centering
    \includegraphics[width=\linewidth]{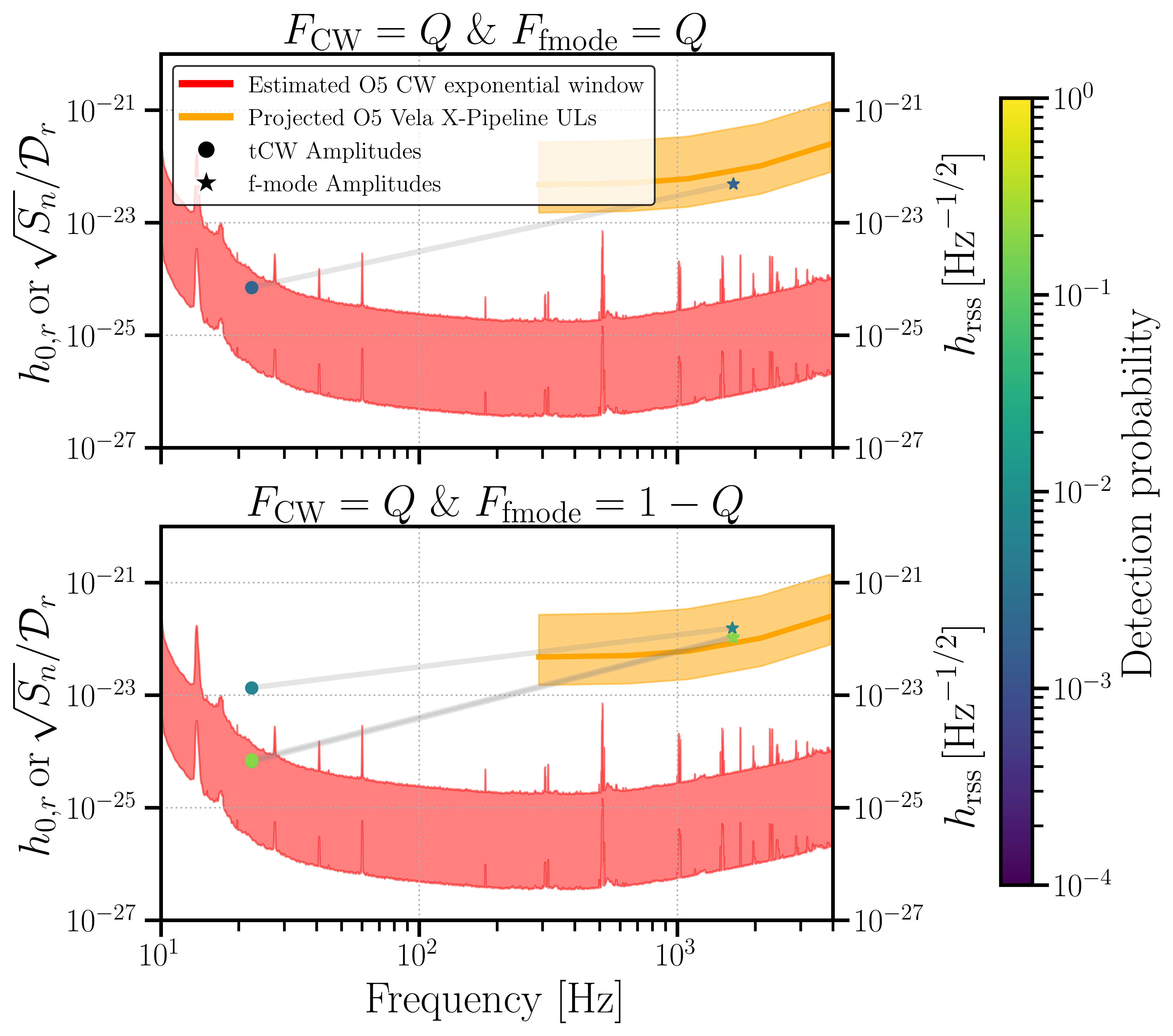}
    \caption{Transient CW and f-mode burst amplitude limits for our catalog of pulsar glitches with measured healing parameters for different values of $\digamma_{\!\mathrm{CW}}$ and $\digamma_{\!\mathrm{fmode}}$ for the LHO + LLO network at O5 sensitivity. Circles show the maximum sampled tCW amplitude $h_{0,\mathrm{CW}}$ and stars show the maximum sampled f-mode amplitude $h_{\mathrm{rss}}$ for each pulsar with a joint-detection probability greater than zero. The points are colored by the joint-detection probability, and gray lines connect tCW points with their corresponding burst point for individual glitches. The red shaded region shows the tCW threshold with the range of values due to different decay times. The orange shaded region shows the $3\sigma$ region for burst upper limits when scaling to different sky locations. }
    \label{fig:O5_detectability}
\end{figure}

\begin{figure}
    \centering
    \includegraphics[width=\linewidth]{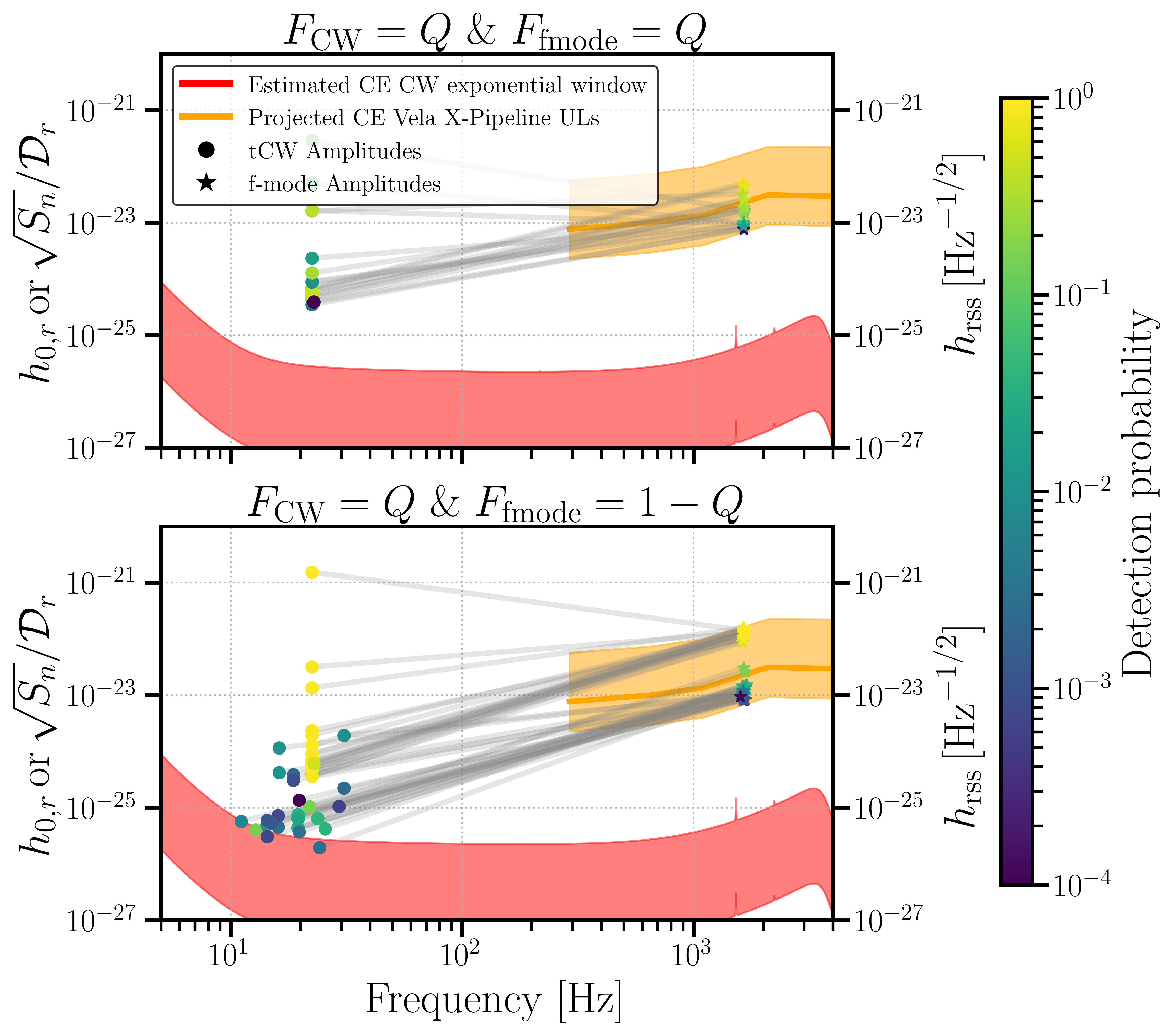}
    \caption{Same as Figure~\ref{fig:O5_detectability} but for a LIGO O5 + Cosmic Explorer network}
    \label{fig:CE_detectability}
\end{figure}

\begin{figure}
    \centering
    \includegraphics[width=\linewidth]{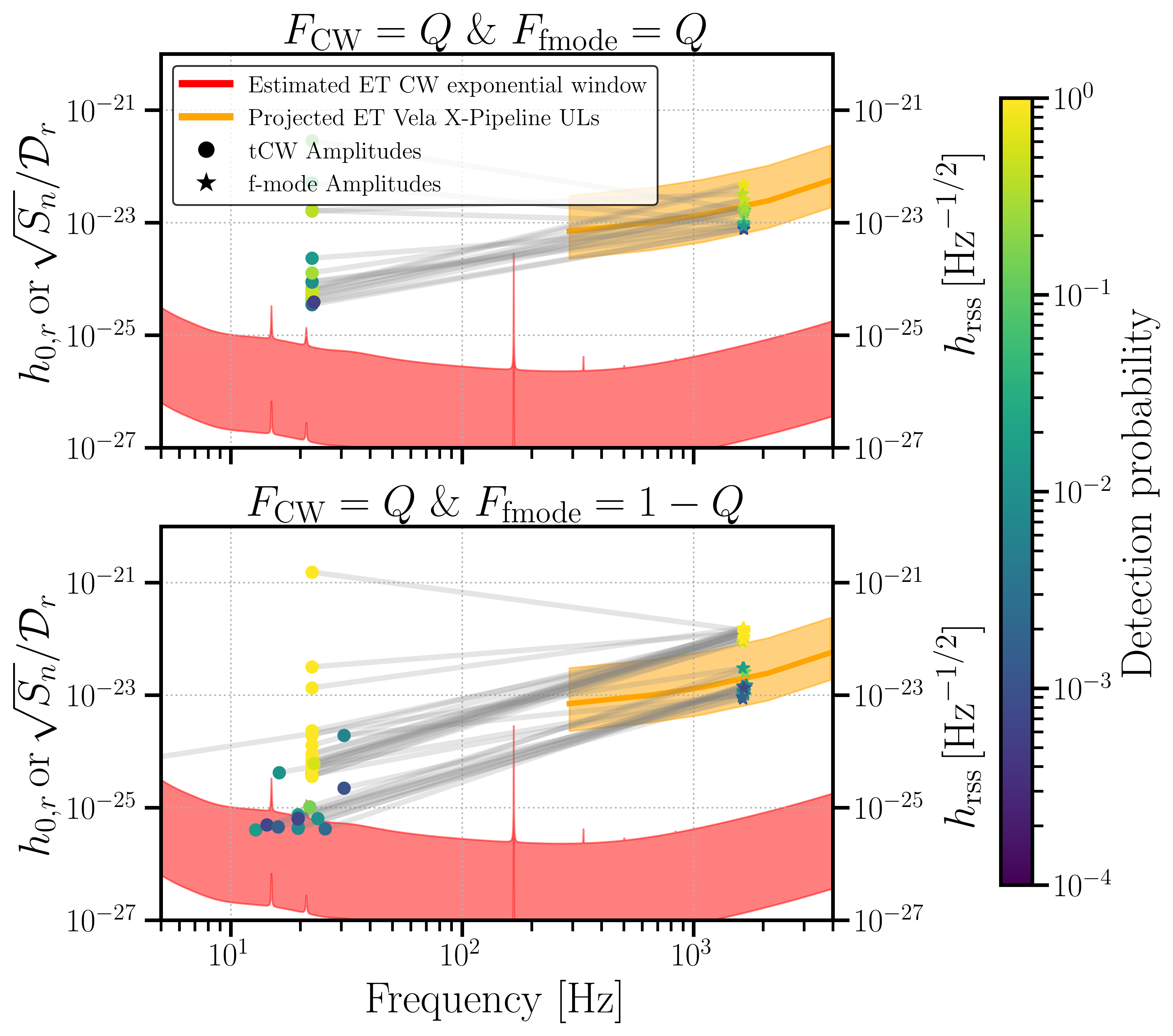}
    \caption{Same as Figure~\ref{fig:O5_detectability} but for a LIGO O5 + Einstein Telescope network}
    \label{fig:ET_detectability}
\end{figure}

\begin{figure}
    \centering
    \includegraphics[width=\linewidth]{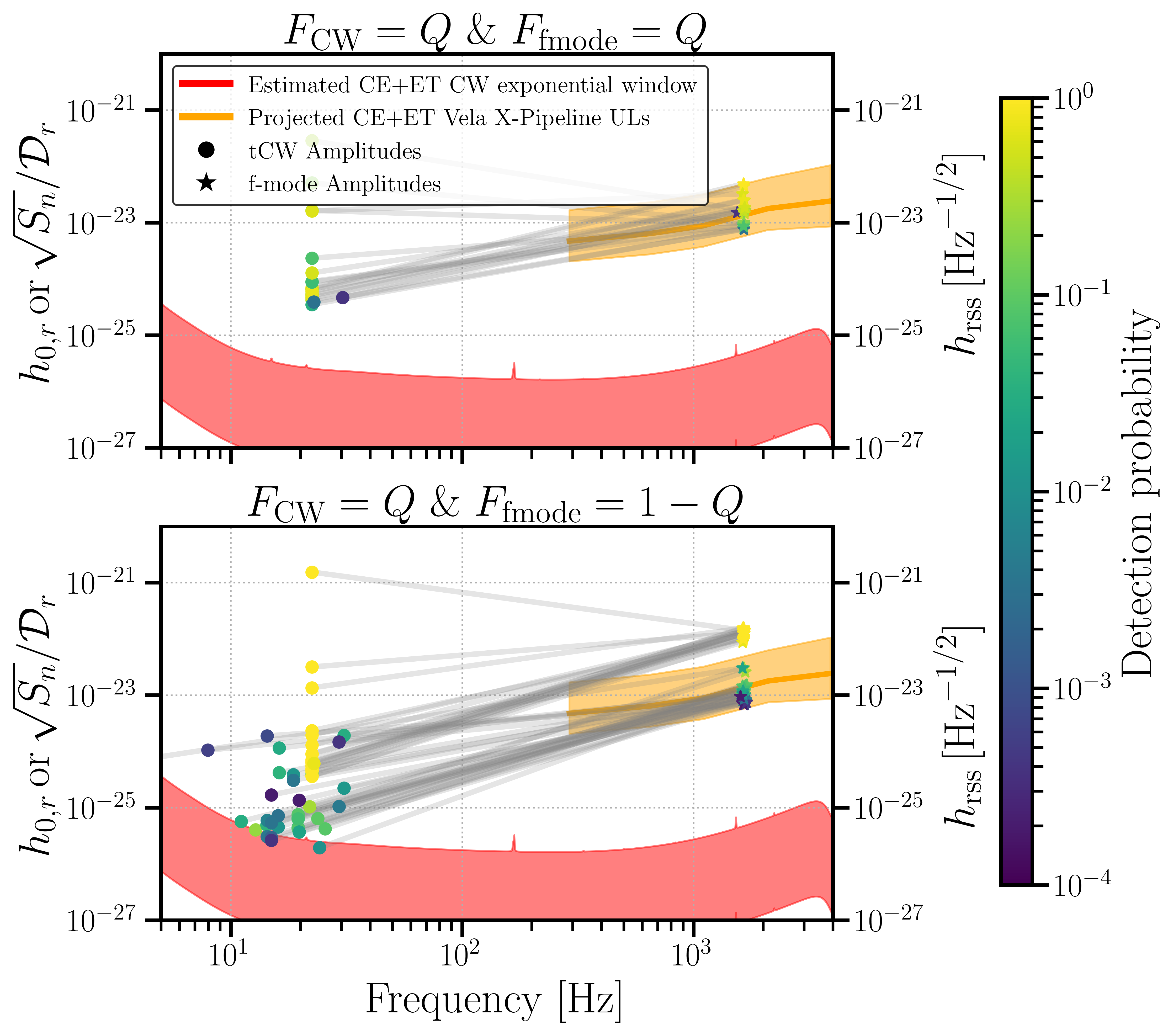}
    \caption{Same as Figure~\ref{fig:O5_detectability} but for a Cosmic Explorer + Einstein Telescope network}
    \label{fig:CE_ET_detectability}
\end{figure}

We compare the Monte Carlo samples drawn for each pulsar to the upper limits estimated for future detectors. Figures~\ref{fig:O5_detectability}, \ref{fig:CE_detectability}, \ref{fig:ET_detectability}, and~\ref{fig:CE_ET_detectability} show the pulsar glitches with at least one Monte Carlo sample identified as ``detectable,'' as described in Section~\ref{sec:methods}, for the O4 network, the O5 network, the Cosmic Explorer \& O5 network, the Einstein Telescope \& O5 network, and the Einstein Telescope \& Cosmic Explorer network, respectively. Approximate detection threshold curves are shown for the tCW and f-mode searches. These are not exact and are intended to help visualize why some glitches are more detectable than others.

To emphasize the degree of effect the different types of searches have on the results, we show both the maximum tCW strain amplitude $h_{0,\mathrm{CW}}$ and the maximum f-mode strain amplitude $h_{\mathrm{rss}}$ for each ``detectable'' pulsar glitch. These points are connected by a gray line for each glitch. We note that in most cases, the tCW amplitude is well above the tCW detectability threshold, but the corresponding f-mode amplitude is barely above the lower end of the spread of the burst threshold.

The spread in the plotted tCW detection thresholds is intended to be representative of the decay time dependence as it is the largest contributing factor to variations in the threshold. The band shown is the network sensitivity (harmonic mean of all detectors in the network) divided by the sensitivity depths calculated in~\cite{moragues_prospects_2023} for a fixed duration of $10\,\mathrm{days}$, and rescaled by a factor of $10\,\mathrm{days}/\tau_{\mathrm{CW}}$. For the threshold shown for the ET + CE network, we reproduce the methodology used in~\cite{moragues_prospects_2023} with the \textsc{COWS3} infrastructure~\cite{Mirasola_2024lcq}. These values are largely consistent with the depths we computed for our thresholds in Section \ref{subsec:CW_ULs}. We emphasize that these are not the thresholds used in the analysis, rather a representation of how much the threshold can vary due to signal parameters.

The spread in the plotted burst detection thresholds is representative of the sky location dependence. We draw samples over a uniform sky location and source orientation and take the $\pm3\sigma$ range of the square root term in Equation~\ref{eq:scale_hrss} applied to the projected $h_{\mathrm{rss}}$ limits for that network to demonstrate the variation in detection threshold. In Figures~\ref{fig:O5_detectability}, \ref{fig:CE_detectability}, \ref{fig:ET_detectability}, and~\ref{fig:CE_ET_detectability}, we plot the maximum amplitude of the f-mode which tends to a common frequency band as detailed in Appendix~\ref{apdx:gw_channel_limits}.

The detectable glitches imply that the tCW analysis is not the limiting factor for a dual detection, rather the sky-location dependence of the burst analysis is generally what may prevent a dual detection. We summarize the results for individual pulsars in the tables in Appendix~\ref{apdx:indiv_pulsars}.

\subsection{Further Exploring the Vela Pulsar}
\label{subsec:probing_vela}

For most of the pulsars, the probability of dual detection is quite low. The main exception to this is the stack of highly detectable glitches with tCW frequency near $22\,\mathrm{Hz}$ corresponding to the Vela pulsar. Our analysis found that for both combinations of values for $\digamma_{\!\mathrm{CW}}$ and $\digamma_{\!\mathrm{fmode}}$, glitches from the Vela pulsar regularly have a high dual-detectability probability at 3rd-generation network sensitivity. Glitches from Vela also are some of the only glitches to have any probability of dual detection at O5 sensitivity. To that end, glitches from the Vela pulsar continue to be high-priority targets in GW astronomy~\cite{10.1093/mnras/stae1659}. Interestingly, we note that glitches from the Crab pulsar were never identified as dual-detectable for any of our scenarios. Despite having a rotation rate placing the expected tCW emission in a more sensitive region of current and future GW detectors near $60\,\mathrm{Hz}$, its smaller glitches (averaging about 2 orders of magnitude smaller than those from the Vela pulsar) greatly diminish the expected f-mode amplitude, leading to this effect.

\begin{figure}
    \centering
    \includegraphics[width=0.9\linewidth]{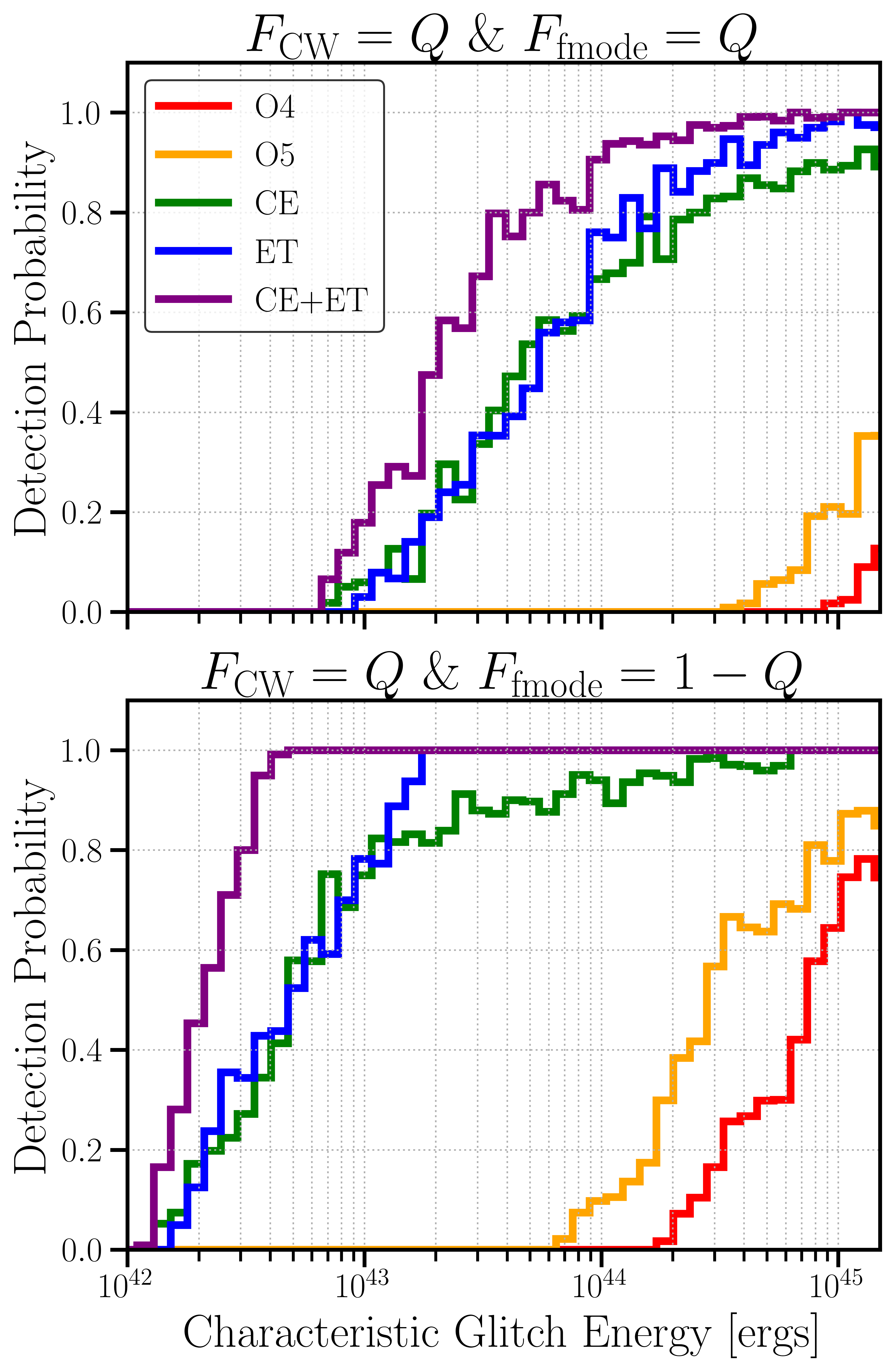}
    \caption{Dual-detectable probabilities for current and future detector networks for the Vela pulsar assuming a mass and radius of $2\,M_{\odot}$ and $15\,\mathrm{km}$, respectively. The healing parameter $Q$ is allowed to vary but is constrained to $Q<0.2$ to match the typical range observed for Vela. Curves show the dual detection probability as a function of the characteristic glitch energy for different energy fractions and detector networks. Samples in discrete energy bins that are considered detectable are summed to estimate the detection probability.}
    \label{fig:vela_detectability}
\end{figure}

We further probe the detectability of the Vela pulsar by exploring the detection probability as a function of the characteristic glitch energy. We draw Monte Carlo samples from similar distributions as described in Section \ref{sec:methods}, but with some changes. We fix the mass and radius to fiducial values of $2\,M_{\odot}$ and $15\,\mathrm{km}$, respectively. For reference, this corresponds to a moment of inertia of about $3.4\times10^{45}\,\mathrm{g\,cm}^2$ (Equation~\ref{eq:inertial_fitting_function}) and an f-mode frequency of about $1600\,\mathrm{Hz}$ (Equations~\ref{eq:fmode_frequency} and~\ref{eq:fmode_rotation}). We use log-uniform distributions for $\digamma_{\!\mathrm{CW}}$, the glitch size $\Delta\nu_s/\nu_s$, and the tCW damping time $\tau_{\mathrm{CW}}$ in order to explore many orders of magnitude. The bounds of these distributions are constructed from the range of reference glitches from Vela. We also constrain the healing parameter $Q$ to values below 0.2 in order to match observed values. For each sample, we compute the tCW and f-mode amplitudes and count the fraction that would be dual-detectable in different characteristic glitch energy (Equation~\ref{eq:glitch_energy}) bins. This is shown in Figure~\ref{fig:vela_detectability}. We can see that while the two networks with at least one 3rd-generation detector will be largely similar in sensitivity, the improvement in detectability over O4 is up to two orders of magnitude in characteristic glitch energy. A network with both of the 3rd-generation detectors sees even greater improvement.

\subsection{Improvement on Constraints in Future Observing Runs}
\label{subsec:3G_constraints}

To estimate what sorts of results may come from future detectors, we test how results would be different if the 2024 Vela glitch had occurred during a 3rd-generation network observing run. We consider the LIGO Hanford and LIGO Livingston detectors at O5 sensitivity with one of the 3rd-generation detectors and a network of two 3rd-generation detectors. Using the methods described in Sections~\ref{subsec:CW_ULs} and \ref{subsec:Burst_ULs}, we can scale the search results from the 2024 Vela glitch analysis \citep{O4_vela_glitch_search} to future detectors and ask how the joint astrophysical implications would change if a non-detection had occurred in a 3rd-generation detector network. As was assumed in~\cite{ball_methods_2025,O4_vela_glitch_search}, we assume GWs were emitted but not detected by the searches.

\subsubsection{Unconstrained Energy Fractions}

\begin{figure}
    \centering
    \includegraphics[width=\linewidth]{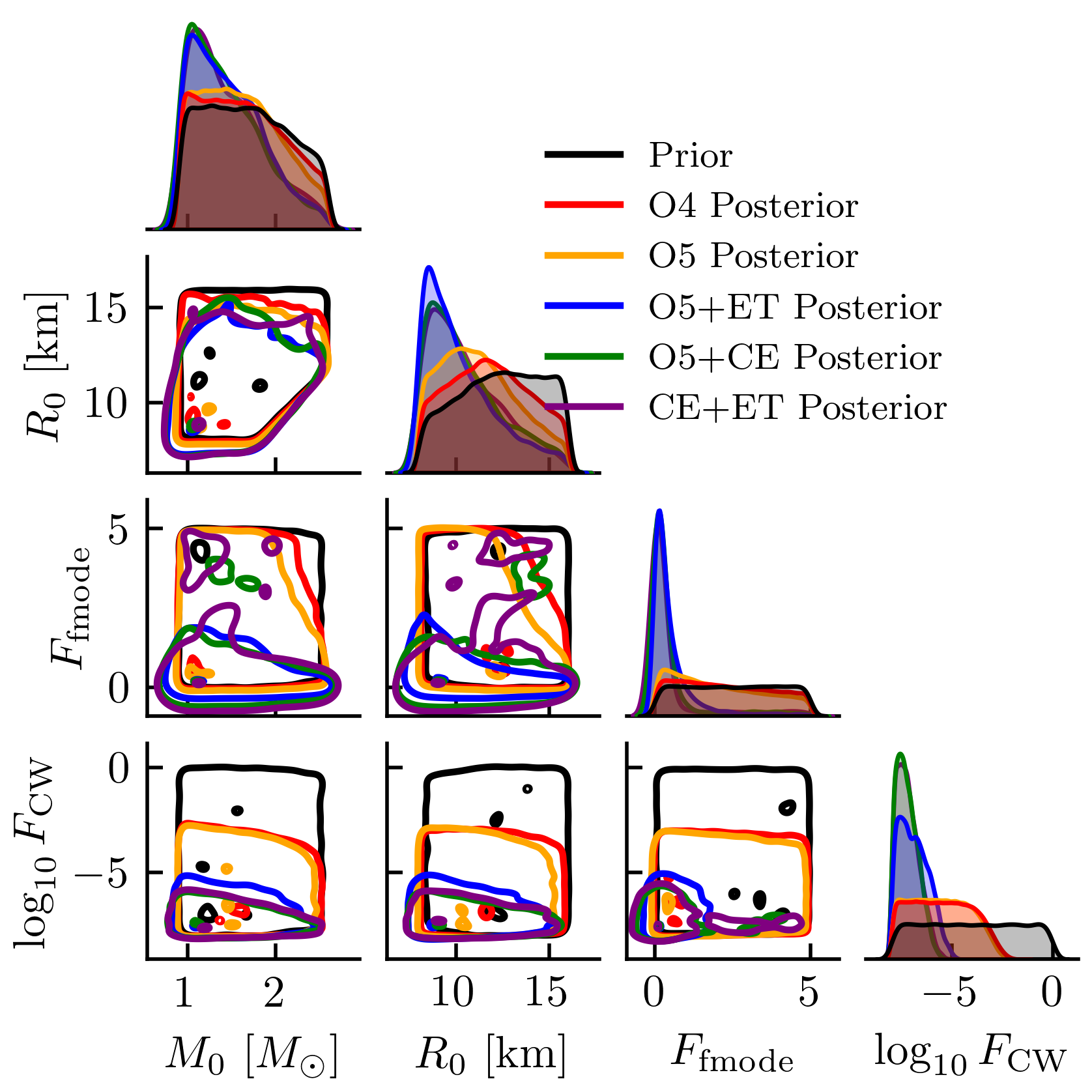}
    \caption{Allowed parameter space for the 2024 Vela pulsar glitch with no detection with limits projected to potential future detector networks. Contours show 10\%, 50\%, and 90\% credible intervals. All future networks place stronger constraints on the energy fractions than the O4 analysis, with 3rd-generation networks being the most constraining.}
    \label{fig:3G_2F_nondetection}
\end{figure}

We first attempt to examine how the constraints on astrophysical parameters change with 3rd-generation detectors when minimal limits are placed on the energy fractions $\digamma_{\!\mathrm{CW}}$ and $\digamma_{\!\mathrm{fmode}}$. This is the same calculation performed in Figure 6 of~\cite{O4_vela_glitch_search}. We expand the prior space for the energy fractions compared to those in~\cite{O4_vela_glitch_search} to $\digamma_{\!\mathrm{CW}}\in[10^{-6},1.0]$ and $\digamma_{\!\mathrm{fmode}}\in[0,5.0]$ to capture the limit of the detectable region. Figure~\ref{fig:3G_2F_nondetection} shows the prior and posterior space for mass, radius, $\digamma_{\!\mathrm{fmode}}$, and $\log_{10}\left(\digamma_{\!\mathrm{CW}}\right)$ for the different configurations. We can see that for all configurations with at least one 3rd-generation detector, limits on the energy fractions are considerably stronger than the constraints at O4 sensitivity. The observed constraints on the mass and radius are correlated with the limits on the energy fraction parameter space.

\subsubsection{Fixed Energy Fractions}

\begin{figure}
    \centering
    \includegraphics[width=\linewidth]{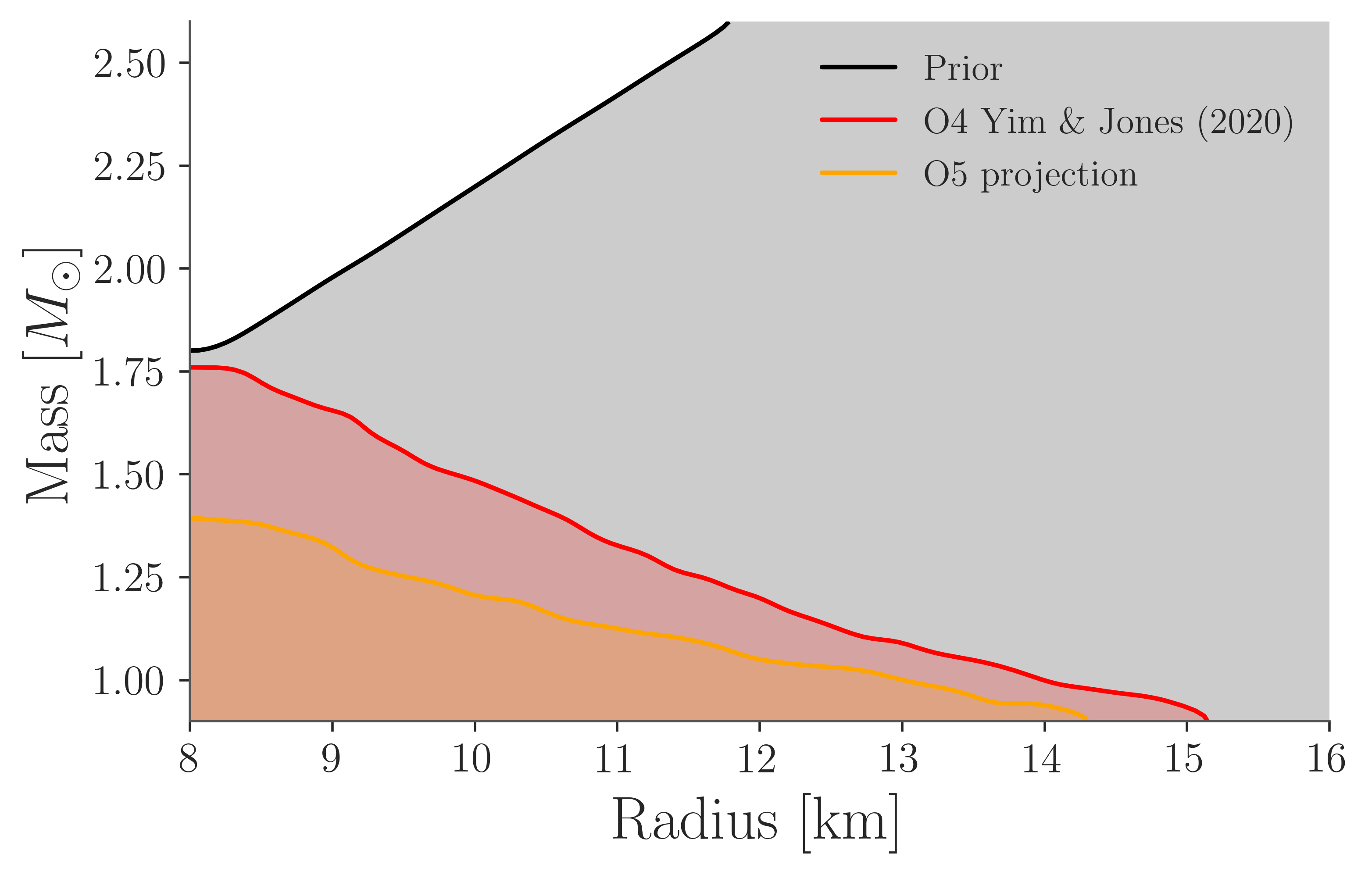}
    \caption{Allowed parameter space for the 2024 Vela pulsar glitch with no detection. Search results from the O4 analysis were projected to O5 sensitivity to estimate how planned upgrades would improve results. We assume a non-detection with fixed $\digamma_{\!\mathrm{CW}}=Q=0.017$ and $\digamma_{\!\mathrm{fmode}}=0.8$ as in \protect\cite{O4_vela_glitch_search}. The red and black contours are reproduced from \protect\cite{O4_vela_glitch_search}, and the orange contour is the allowed parameter space from a non-detection from the LIGO detectors at O5 sensitivity using the updated universal relation for moment of inertia (Equation~\ref{eq:inertial_fitting_function}). Under this non-detection scenario with the \protect\cite{yim_transient_2020} model, O5 sensitivity improves constraints by more than a quarter of a solar mass and nearly $1\,\mathrm{km}$ in radius. The 3G detector networks were also tested, but none of the prior parameter space was permitted in any permutation.}
    \label{fig:3G_nondetection}
\end{figure}

We now aim to repeat the calculation performed for Figure 7 in~\cite{O4_vela_glitch_search}. This analysis fixed the energy fractions to $\digamma_{\!\mathrm{CW}}=Q=0.017$ from the~\cite{yim_transient_2020} model and $\digamma_{\!\mathrm{fmode}}=0.8$ and computed what masses and radii would be permitted under the assumption that GWs in both channels were emitted but not detected. We use the same prior distributions as them for our analysis here. In Figure~\ref{fig:3G_nondetection} we show the results for repeating this with search results projected to LIGO O5 sensitivity. Had the LIGO detectors been at O5 sensitivity, constraints on the mass and radius would be improved by more than a quarter of a solar mass and nearly $1\,\mathrm{km}$, respectively. We note that networks with 3rd-generation detectors sensitivity resulted in zero allowed mass/radius values. These results suggest that a non-detection of a glitch similar to the 2024 Vela glitch at 3rd-generation sensitivity would likely not permit the model in~\cite{yim_transient_2020} and could potentially act as an argument to disprove the model entirely.

\begin{figure*}
    \centering
    \begin{subfigure}[t]{0.33\textwidth}
        \centering
        \includegraphics[width = \linewidth]{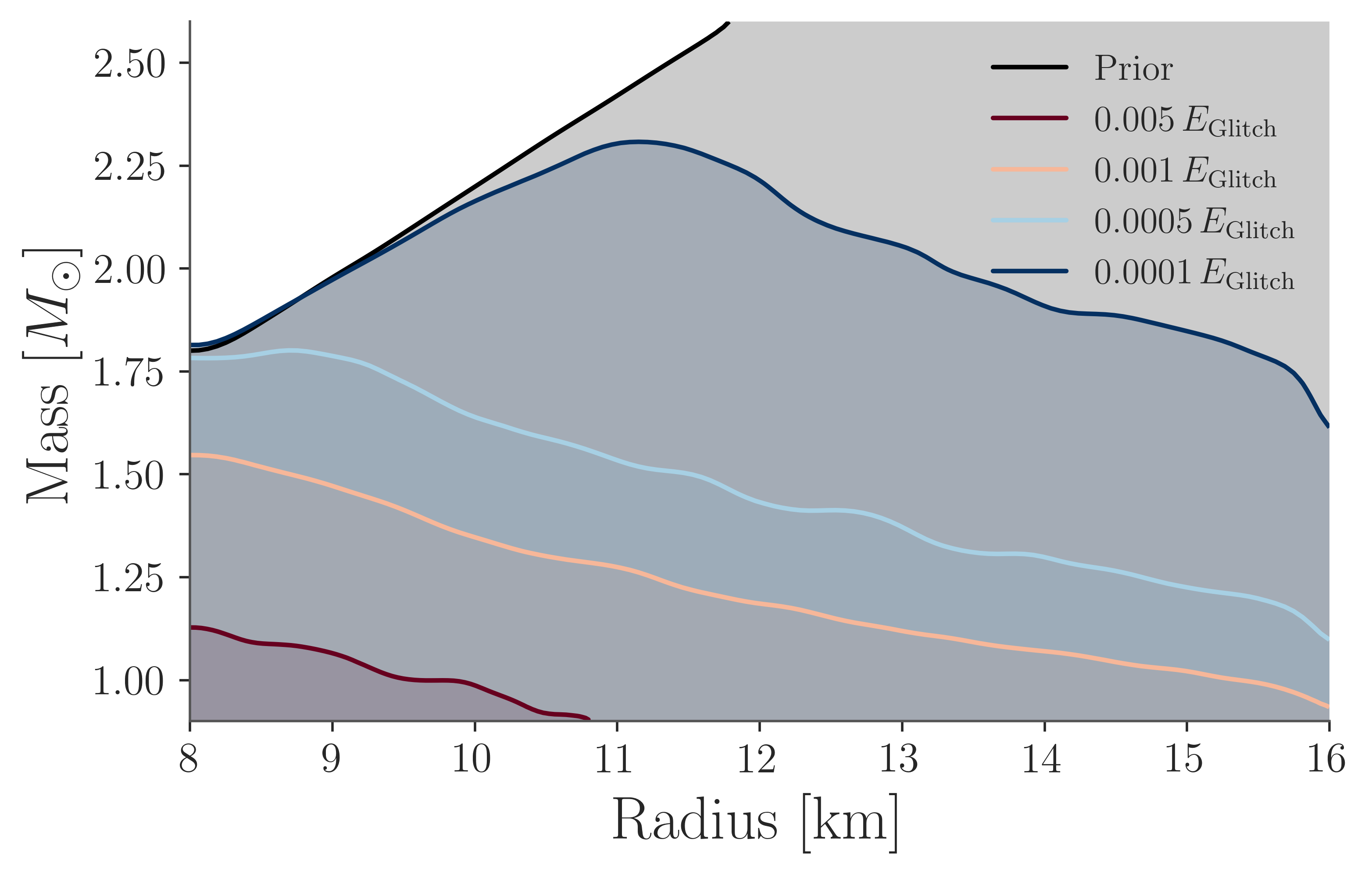}
        \caption{O5 with Einstein Telescope}
        \label{fig:ET_nondetection_pessimistic}
    \end{subfigure}%
    ~ 
    \begin{subfigure}[t]{0.33\textwidth}
        \centering
        \includegraphics[width = \linewidth]{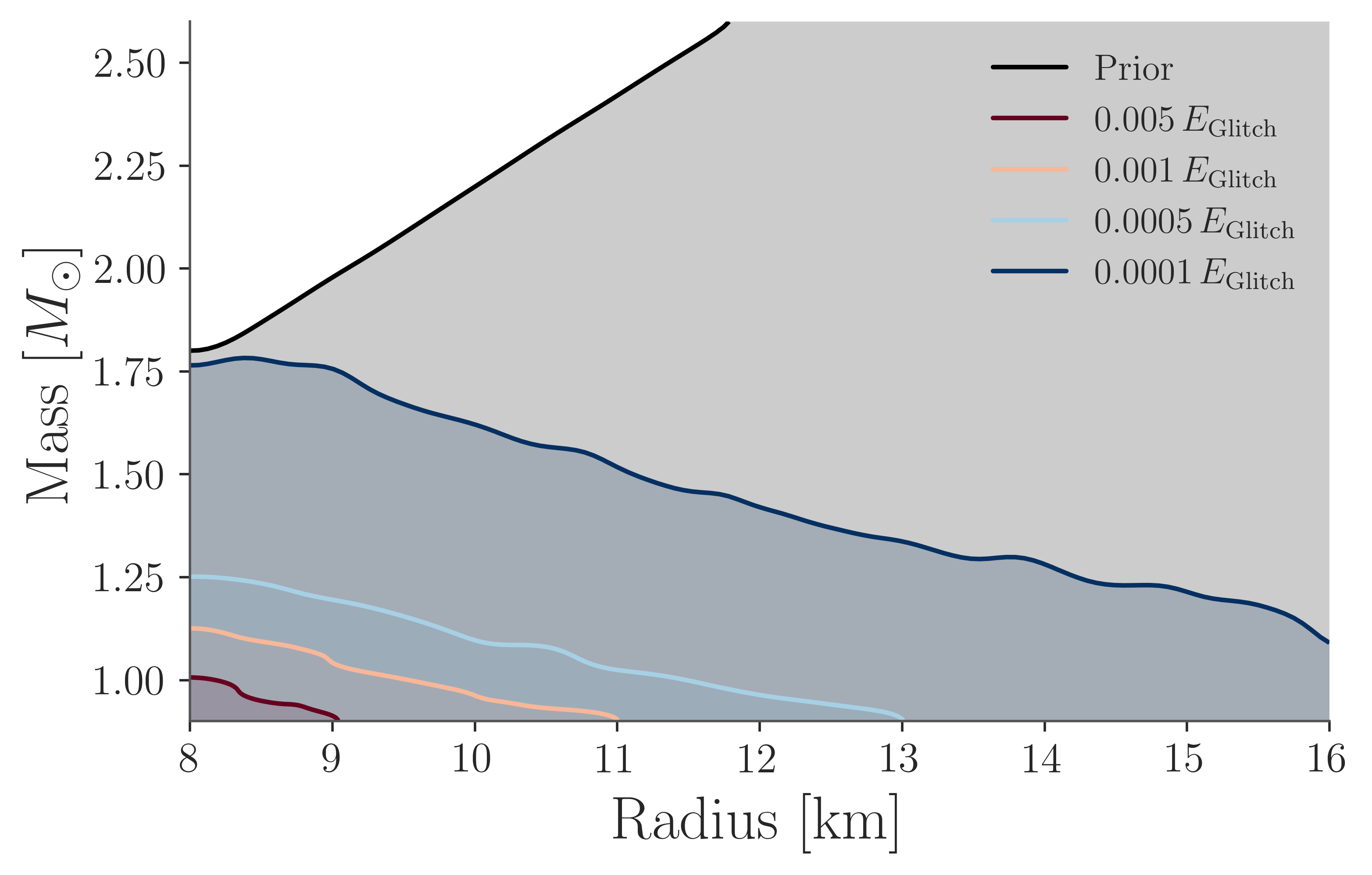}
        \caption{O5 with Cosmic Explorer}
        \label{fig:CE_nondetection_pessimistic}
    \end{subfigure}%
    ~
    \begin{subfigure}[t]{0.33\textwidth}
        \centering
        \includegraphics[width = \linewidth]{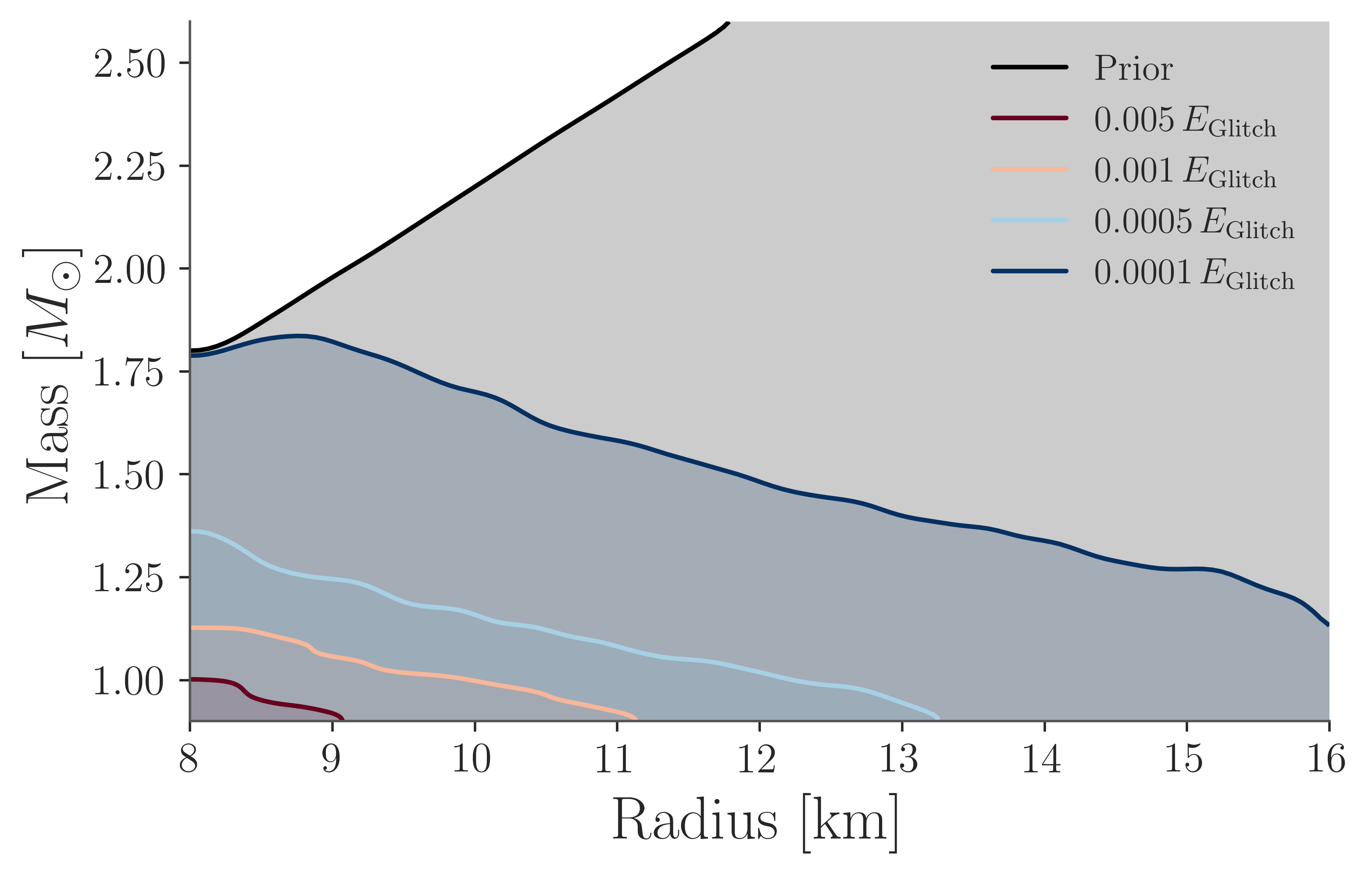}
        \caption{Cosmic Explorer and the Einstein Telescope}
        \label{fig:CE_ET_nondetection_pessimistic}
    \end{subfigure}
    \caption{Allowed parameter space for the 2024 Vela pulsar glitch in future detector networks assuming both tCW and burst signals were emitted but not detected by tCW or burst searches. Search results from the O4 analysis were projected to different hypothetical 3G detector networks. We assume a non-detection with fixed energy fractions but with a global factor $\digamma_{\!\mathrm{tot}}$ limiting the total GW energy such that $\digamma_{\!\mathrm{CW}}+\digamma_{\!\mathrm{fmode}} = \digamma_{\!\mathrm{tot}}$. We then have $\digamma_{\!\mathrm{CW}}/\digamma_{\!\mathrm{tot}}=Q=0.017$ and $\digamma_{\!\mathrm{fmode}}/\digamma_{\!\mathrm{tot}}=1-Q$. Contours show the allowed parameter space assuming different total GW energies relative to the characteristic glitch energy. Any of the 3G network would make strong statements from a non-detection even if the total GW energy emitted was a small fraction of the characteristic glitch energy. Likely due to improved sensitivity near $22\,\mathrm{Hz}$ and thus improved tCW limits, the O5 \& Cosmic Explorer network would place stronger constraints than the O5 \& Einstein Telescope network. The best limits would be reached by a network of Cosmic Explorer and the Einstein Telescope.}
\end{figure*}

To better understand the limits that may be set by next-generation detectors, we now consider the case where the total GW energy is much less than the characteristic glitch energy (Equation~\ref{eq:glitch_energy}). We introduce a global factor $\digamma_{\!\mathrm{tot}}$ modulating the total GW energy emitted by the glitch as a fraction of the characteristic glitch energy such that $\digamma_{\!\mathrm{CW}}+\digamma_{\!\mathrm{fmode}}=\digamma_{\!\mathrm{tot}}$. With this energy budget, we fix $\digamma_{\!\mathrm{CW}}/\digamma_{\!\mathrm{tot}}=Q=0.017$ and $\digamma_{\!\mathrm{fmode}}/\digamma_{\!\mathrm{tot}}=1-Q$. We decrease this total energy fraction to see how a hypothetical 3rd-generation GW detector network would perform had it been online during the 2024 Vela glitch. Figure~\ref{fig:ET_nondetection_pessimistic} shows the range of allowed parameters assuming a non-detection by a LIGO O5 and Einstein Telescope network, Figure~\ref{fig:CE_nondetection_pessimistic} shows the range of allowed parameters assuming a non-detection by a LIGO O5 and Cosmic Explorer network, and Figure~\ref{fig:CE_ET_nondetection_pessimistic} shows the range of allowed parameters assuming a non-detection by a Cosmic Explorer and Einstein Telescope network. The contours correspond to different $\digamma_{\!\mathrm{tot}}$ expressed as fractions of the characteristic glitch energy $E_{\mathrm{Glitch}}$. We can see that unless we assume the total GW energy is a very small fraction of the characteristic glitch energy, strong constraints on the pulsar mass and radius can be placed from a non-detection. An O5 + ET network would approximately match the constraints of the O4 network assuming only $\sim 0.05\%$ to $\sim 0.1\%$ of the characteristic energy was emitted as GWs. An O5 + CE network and a CE + ET network would only need to assume $\sim0.01\%$ to $\sim0.05\%$.

\section{Conclusions}
\label{sec:conclusions}
We have calculated the probability of simultaneously detecting long-duration tCW and burst-like f-mode GWs from pulsar glitches in future detector configurations. We projected sensitivity estimates from O4 to O5, Cosmic Explorer, and the Einstein Telescope in order to calculate the probability of dual detection for a set of pulsar glitches from the ATNF catalog. This analysis was performed assuming the transient CW energy budget predicted by~\cite{yim_transient_2020} and two different potential energy budgets for the f-mode.

We have shown that, perhaps obviously, next-generation GW detectors like Cosmic Explorer and the Einstein Telescope will greatly improve the chances of making a dual detection. However, even with the improved sensitivity, these detections are not guaranteed. Limited primarily by the burst analyses, dual detection is highly dependent on the sky location of the source pulsar and the size of the glitch. Even so, we can say that glitches from the Vela pulsar are among the strongest candidates for future dual detections.

By projecting previous non-detection upper limits to future detectors, we have shown that had the 2024 Vela pulsar glitch occurred during a 3rd-generation observing run, it very likely would have been detectable by both tCW and burst searches, and a non-detection could have potentially ruled out the~\cite{yim_transient_2020} model for tCW emission entirely.

These results demonstrate the improvement that next-generation detectors will bring to the analysis of pulsar glitches. It is likely that both continuous and burst gravitational waves from a single pulsar glitch could be measured with the improved sensitivity. Such a measurement could have profound implications for the understanding of the mechanisms behind pulsar glitches as well as the underlying neutron star equation of state. In turn, a double non-detection would provide strong constraints on the emission models used here.

Future work could explore how different priors informed by other astrophysical measurements of neutron stars may impact the results here.

\begin{acknowledgments}

The authors would like to thank Shubhanshu Tiwari for his useful comments and discussions.

This material is based upon work supported by NSF's LIGO Laboratory which is a major facility fully funded by the National Science Foundation.

This work was supported by the Universitat de les Illes Balears (UIB) with funds from the Programa de Foment de la Recerca i la Innovació de la UIB 2024-2026 (supported by the yearly plan of the Tourist Stay Tax ITS2023-086); the Spanish Agencia Estatal de Investigación grants PID2022-138626NB-I00, RED2024-153978-E, RED2024-153735-E, funded by MICIU/AEI/10.13039/501100011033 and the ERDF/EU; and the Comunitat Autònoma de les Illes Balears through an ``Ajut per a projectes de recerca científica i tecnològica'' (PRD2025\_100) and through the Conselleria d'Educació i Universitats with funds from the European Union - European Regional Development Fund (ERDF) (SINCO2022/18146 - Plataforma HiTech-IAC3-BIO); and by COST action SCALES CA24139, supported by COST (European Cooperation in Science and Technology).

This paper has been assigned LIGO DCC number LIGO-P2600533 and Einstein Telescope document number ET-0768B-26.

\end{acknowledgments}

\appendix

\section{Understanding the GW Universal Relations}
\label{apdx:gw_channel_limits}
\begin{figure*}
    \centering
    \begin{subfigure}[t]{0.33\textwidth}
        \centering
        \includegraphics[width = \linewidth]{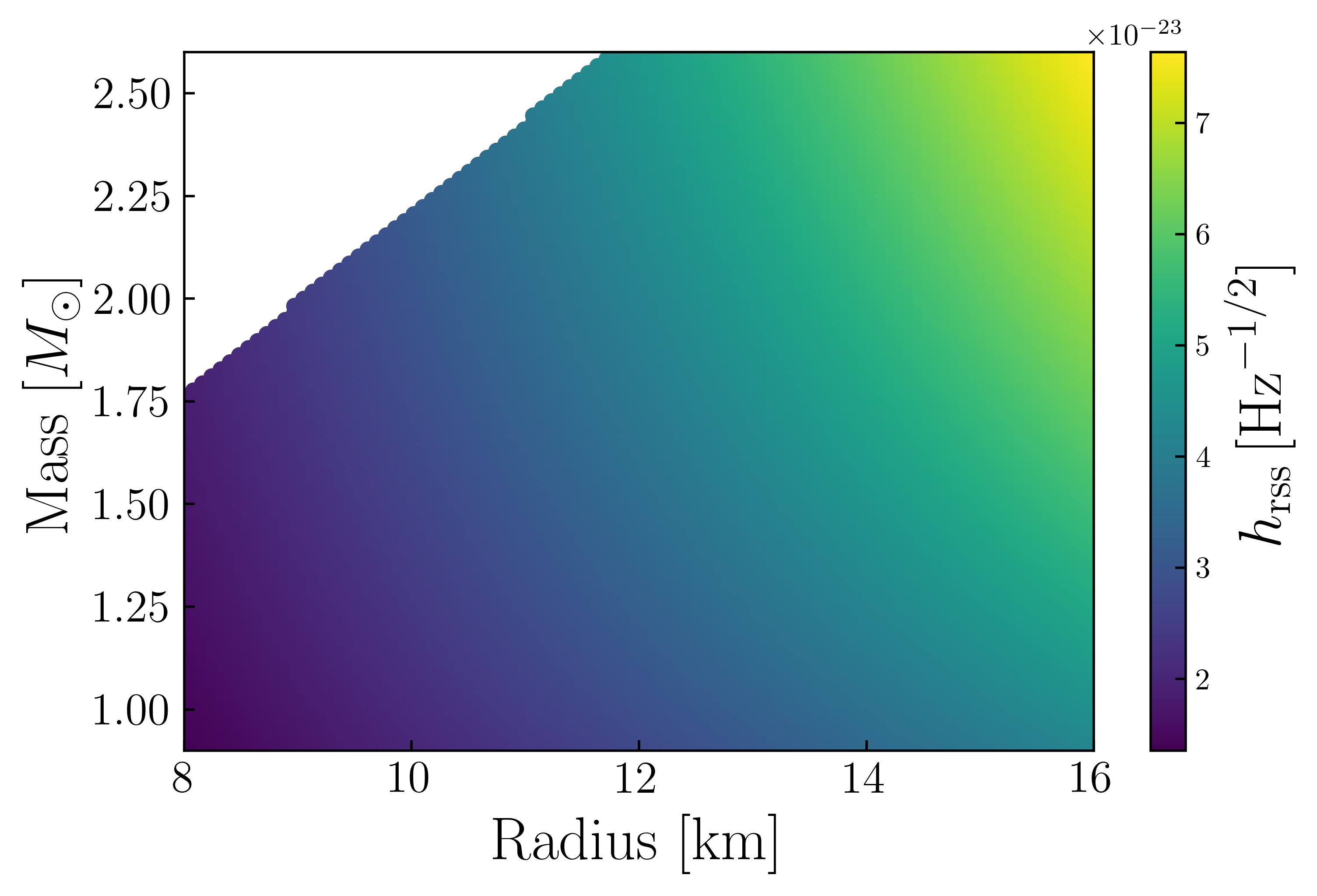}
        \caption{}
        \label{fig:fmode_MR_hrss}
    \end{subfigure}%
    ~ 
    \begin{subfigure}[t]{0.33\textwidth}
        \centering
        \includegraphics[width = \linewidth]{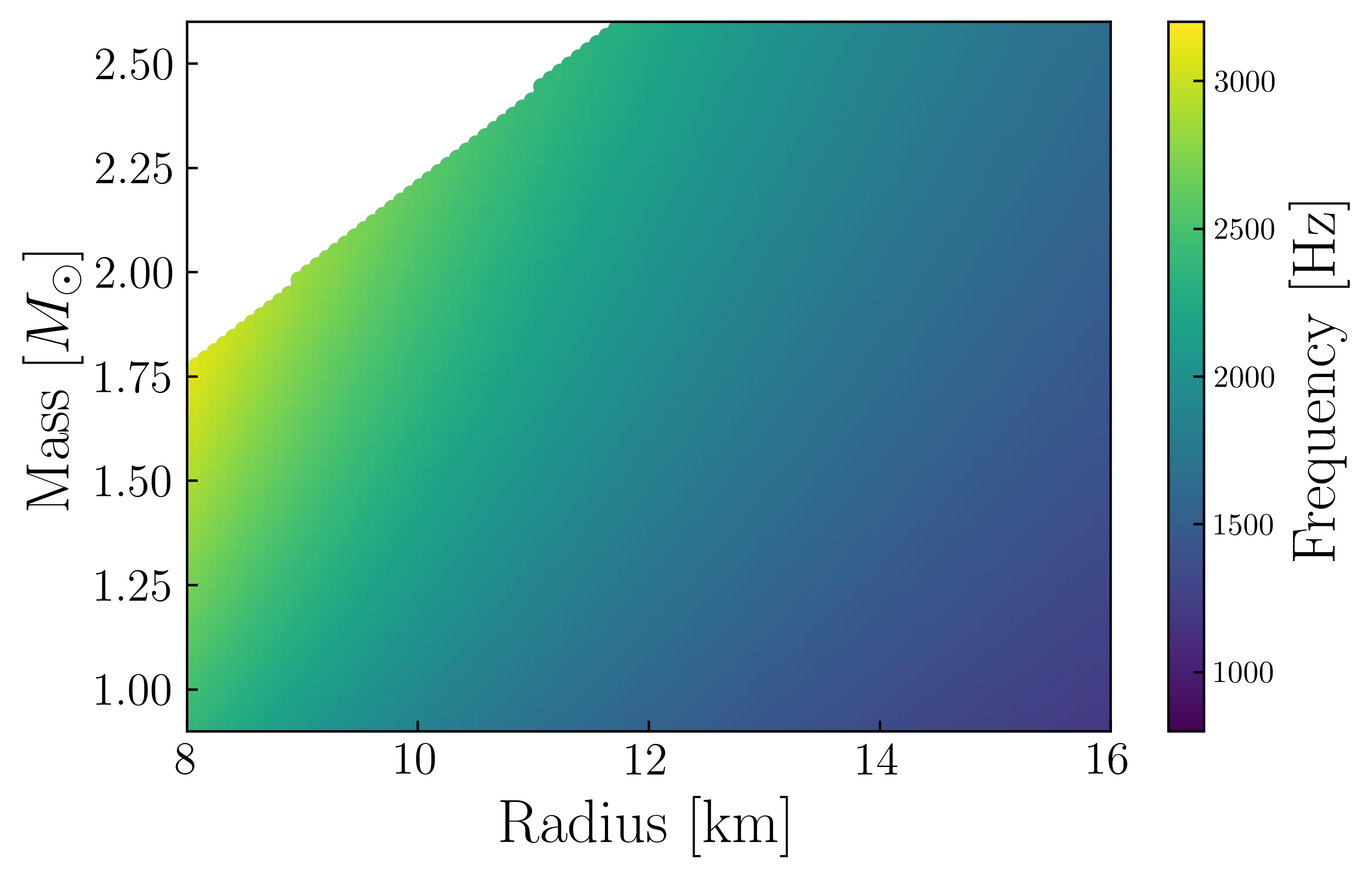}
        \caption{}
        \label{fig:fmode_MR_freq}
    \end{subfigure}%
    ~ 
    \begin{subfigure}[t]{0.33\textwidth}
        \centering
        \includegraphics[width = \linewidth]{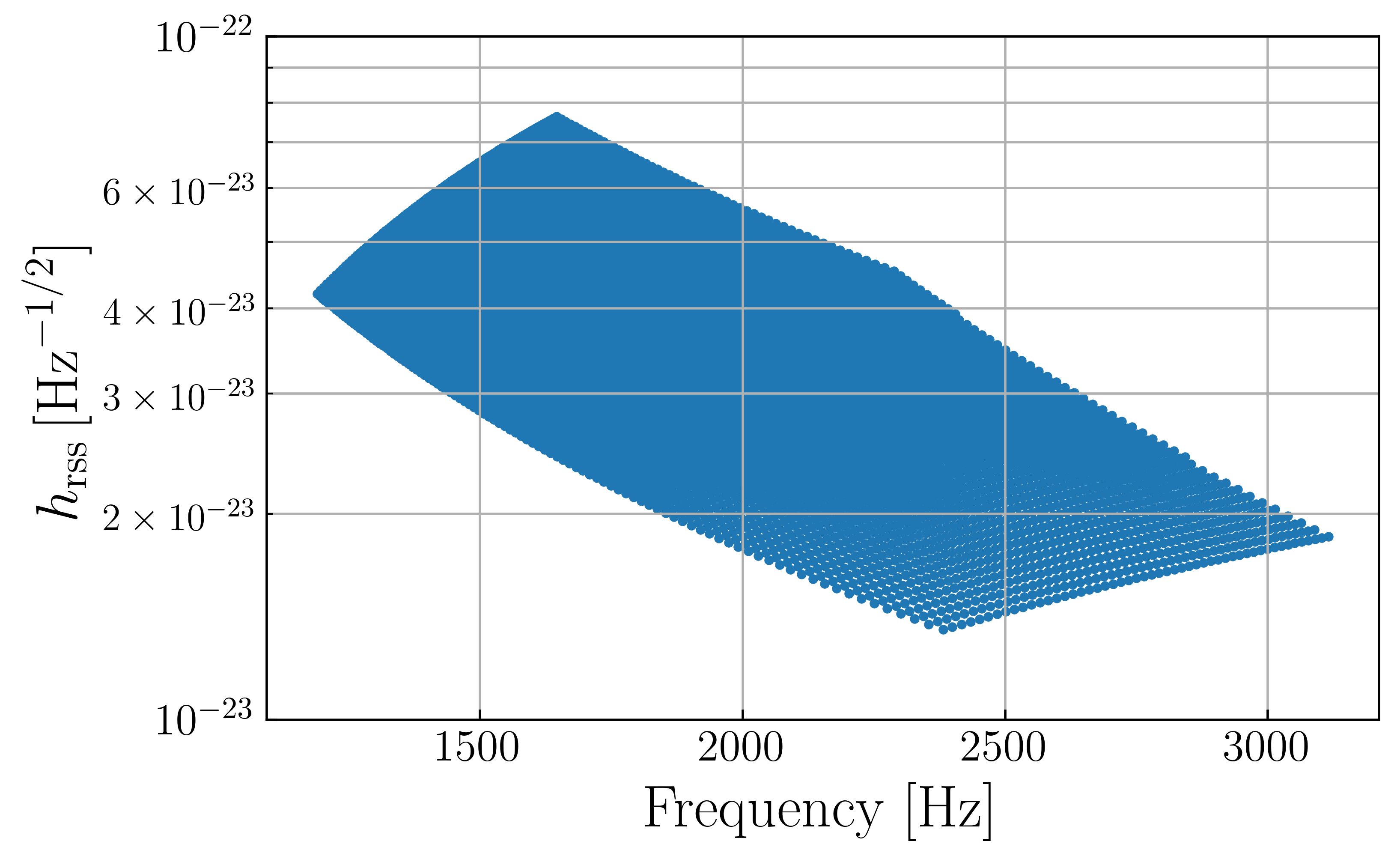}
        \caption{}
        \label{fig:fmode_MR_freq}
    \end{subfigure}

    \caption{Parameters of the f-mode model showing why the peak f-mode $h_{\mathrm{rss}}$ does not occur at a maximum or minimum frequency. \textbf{(a)} plots the $h_{\mathrm{rss}}$ as a function of mass and radius, which shows that $h_{\mathrm{rss}}$ is maximized at high mass and radius. \textbf{(b)} plots the $\nu_f$ as a function of mass and radius, which shows that frequency is maximized at low radius and high mass but is minimized at high radius and low mass. \textbf{(c)} plots the permitted values of $h_{\mathrm{rss}}$ against $\nu_f$ for the allowed mass/radius space, which shows how the $h_{\mathrm{rss}}$ is maximized at an intermediate frequency, near $1650\,\mathrm{Hz}$.}
    \label{fig:apdx:fmode_hrss_frequency}
\end{figure*}

The $h_{\mathrm{rss}}$ for an f-mode is maximized at an unintuitive frequency in our model (as may be evident in the f-mode frequencies of peak $h_{\mathrm{rss}}$ in Figures~\ref{fig:O5_detectability}, \ref{fig:CE_detectability}, \ref{fig:ET_detectability}, and \ref{fig:CE_ET_detectability}). Figure~\ref{fig:apdx:fmode_hrss_frequency} shows that the $h_{\mathrm{rss}}$ is maximized at high mass and high radius values with slightly stronger dependence on radius than mass, but the frequency is maximized at low radius and high mass values. The exact values used for these plots use a uniform grid of mass and radius values with the same spin as the Vela pulsar and glitch size as its 2024 glitch. This results in the frequency associated with the maximum $h_{\mathrm{rss}}$ being near $1650\,\mathrm{Hz}$. The missing corner at low radius and high mass is due to removing all mass/radius points that violate causality or would produce a black hole. The jagged edge is a plotting artifact due to the mass/radius grid.

\begin{figure*}
    \centering
    \begin{subfigure}[t]{0.48\textwidth}
        \centering
        \includegraphics[width=0.87\linewidth]{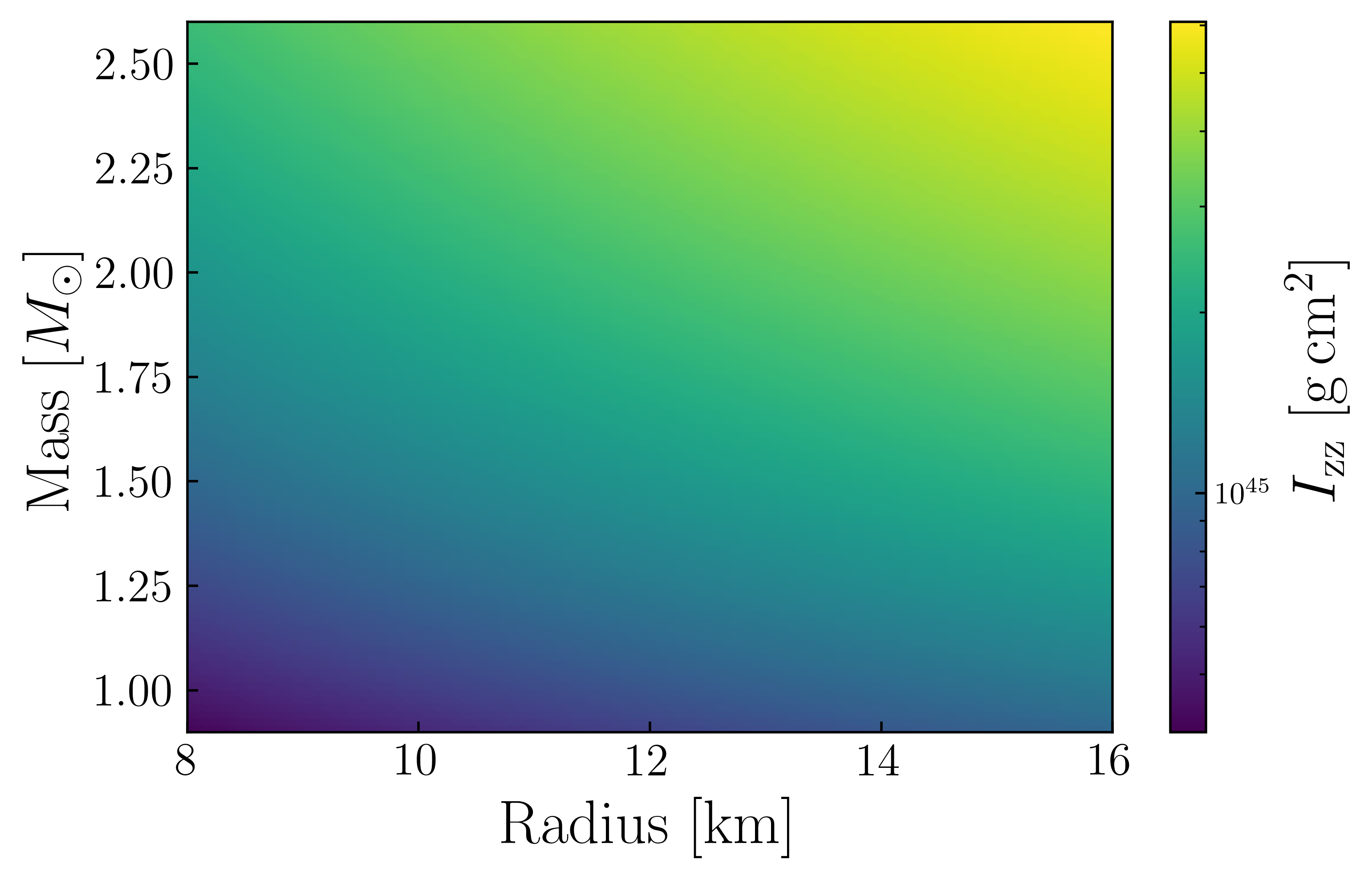}
        \caption{}
        \label{fig:inertia_function}
    \end{subfigure}%
    ~
    \begin{subfigure}[t]{0.48\textwidth}
        \centering
        \includegraphics[width=0.9\linewidth]{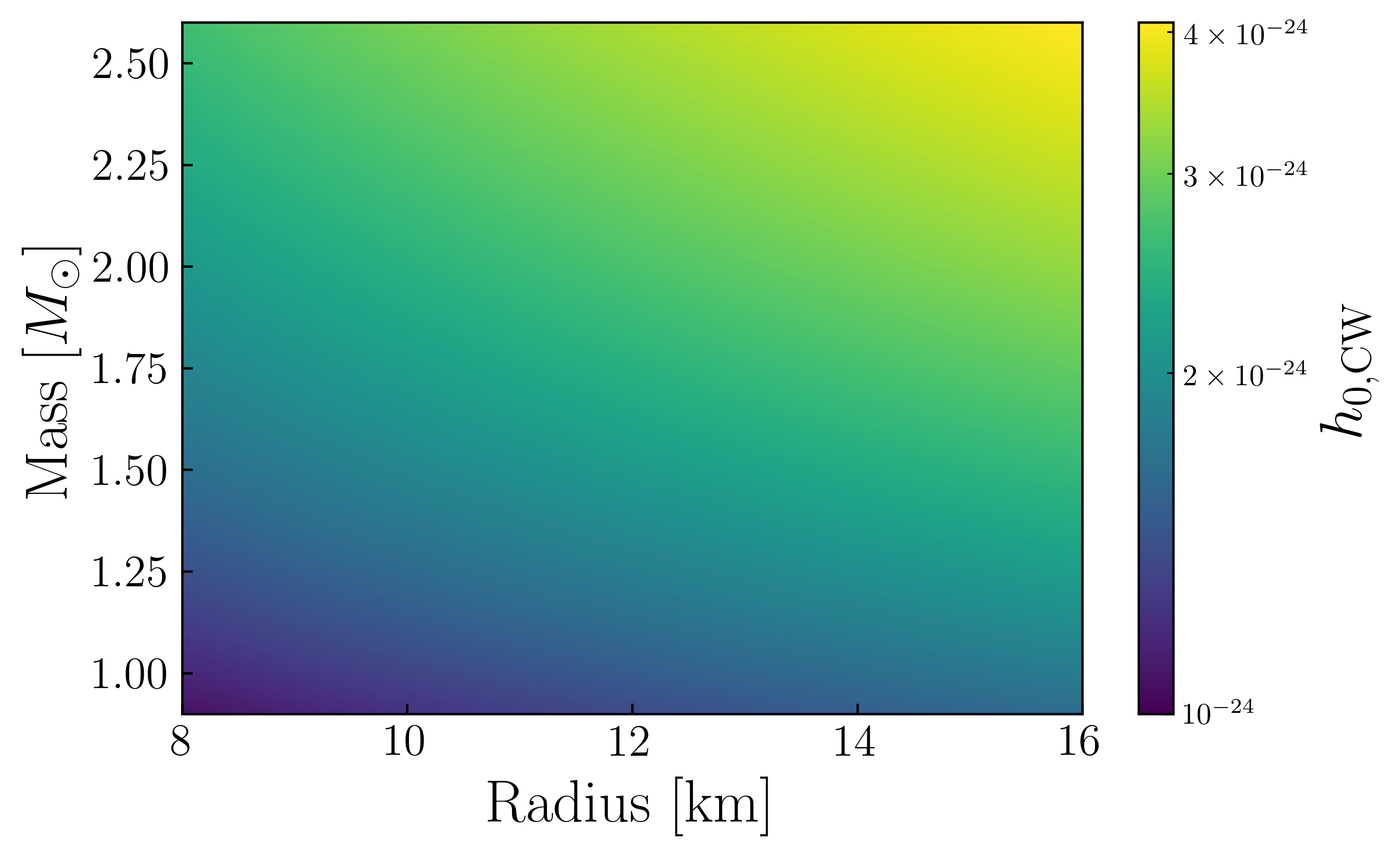}
        \caption{}
        \label{fig:CW_h0_vs_MR}
    \end{subfigure}
    \caption{Mass and radius dependence of \textbf{(a)} the moment of inertia relation for a neutron star and \textbf{(b)} the tCW amplitude $h_{0,\mathrm{CW}}$ for the 2024 Vela pulsar glitch. The mass and radius dependence of $h_{0,\mathrm{CW}}$ is exclusively driven by the moment of inertia relation. In both cases, the values maximize at high mass and high radius.}
\end{figure*}

We can use a similar procedure to examine how the pulsar mass and radius affect the amplitude of the tCW. Figure~\ref{fig:inertia_function} shows the relationship between neutron star mass and radius and the moment of inertia from Equation~\ref{eq:inertial_fitting_function} and Figure~\ref{fig:CW_h0_vs_MR} shows how the tCW amplitude $h_{0,\mathrm{CW}}$ is dependent on mass and radius. These two plots are nearly identical since all mass/radius dependence on $h_{0,\mathrm{CW}}$ comes from the moment of inertia. These parameters are maximized by high mass and high radius, similar to the f-mode $h_{\mathrm{rss}}$, but not quite as dependent on the radius.

\section{Pulsar specific sensitivity}
\label{apdx:indiv_pulsars}
We now consider glitches grouped by individual pulsars rather than all glitches individually. The following tables show the maximum $h_{0,\mathrm{CW}}$ and $h_{\mathrm{rss}}$ values from all samples from all glitches for each pulsar. This represents a reasonable maximum GW amplitude from each pulsar. The number of glitches with at least one sample above both the burst and tCW detection thresholds are shown for different detector configurations in the following tables. We emphasize that just because a pulsar has a non-zero detectability percentage, it is not necessarily a high percentage. A ``detectable'' pulsar could have had as few as $1/10000$ samples meeting the threshold for detection in both GW channels.

\begin{table*}
\centering
\caption{Per-pulsar detection percentage with energy fractions $\digamma_{\!\mathrm{CW}}=Q$ and $\digamma_{\!\mathrm{fmode}}=1-Q$ such that they sum to the total glitch energy. The maximum GW amplitudes are shown for each pulsar, maximizing across all glitches and samples for each glitch. The number of glitches with at least one sample above both CW and Burst detection thresholds are shown for the different detector configurations. Empty positions correspond to pulsars where the only available glitch had $Q>1$ which was incompatible with the energy assumption here.}
\label{tab:CW_Q_fmode_1-Q}
\begin{tabular}{ccccrrrrr}
\hline
    &    PSR     &  Max $h_{0,\mathrm{CW}}$  &  Max $h_{\mathrm{rss}}\,[\mathrm{Hz}^{-1/2}]$  &      O4      &      O5      &    O5 + CE    &    O5 + ET    &    ET + CE    \\
\hline
 1  & J0146+6145 &     -      &     -      &  0/1 (0 \%)  &  0/1 (0 \%)  &   0/1 (0 \%)   &   0/1 (0 \%)   &   0/1 (0 \%)   \\
 2  & J0205+6449 &  3.6e-25   &  1.1e-23   &  0/1 (0 \%)  &  0/1 (0 \%)  &   0/1 (0 \%)   &   0/1 (0 \%)   &   0/1 (0 \%)   \\
 3  & J0358+5413 &  4.0e-26   &  3.1e-23   &  0/1 (0 \%)  &  0/1 (0 \%)  &  1/1 (100 \%)  &  1/1 (100 \%)  &  1/1 (100 \%)  \\
 4  & J0528+2200 &  1.6e-24   &  1.9e-26   &  0/2 (0 \%)  &  0/2 (0 \%)  &   0/2 (0 \%)   &   0/2 (0 \%)   &   0/2 (0 \%)   \\
 5  & J0534+2200 &  1.0e-23   &  2.6e-24   & 0/10 (0 \%)  & 0/10 (0 \%)  &  0/10 (0 \%)   &  0/10 (0 \%)   &  0/10 (0 \%)   \\
 6  & J0631+1036 &  6.5e-25   &  7.4e-25   &  0/2 (0 \%)  &  0/2 (0 \%)  &   0/2 (0 \%)   &   0/2 (0 \%)   &   0/2 (0 \%)   \\
 7  & J0729-1448 &  1.1e-24   &  8.9e-24   &  0/1 (0 \%)  &  0/1 (0 \%)  &   0/1 (0 \%)   &   0/1 (0 \%)   &  1/1 (100 \%)  \\
 8  & J0835-4510 &  1.5e-21   &  1.6e-22   & 2/19 (11 \%) & 3/19 (16 \%) & 19/19 (100 \%) & 19/19 (100 \%) & 19/19 (100 \%) \\
 9  & J0940-5428 &  6.1e-25   &  8.7e-23   &  0/1 (0 \%)  &  0/1 (0 \%)  &  1/1 (100 \%)  &  1/1 (100 \%)  &  1/1 (100 \%)  \\
 10 & J1015-5719 &  5.0e-26   &  1.1e-23   &  0/1 (0 \%)  &  0/1 (0 \%)  &  1/1 (100 \%)  &  1/1 (100 \%)  &  1/1 (100 \%)  \\
 11 & J1016-5857 &  3.8e-25   &  9.2e-24   &  0/2 (0 \%)  &  0/2 (0 \%)  &  2/2 (100 \%)  &   0/2 (0 \%)   &  2/2 (100 \%)  \\
 12 & J1028-5819 &  1.0e-25   &  2.6e-23   &  0/1 (0 \%)  &  0/1 (0 \%)  &  1/1 (100 \%)  &  1/1 (100 \%)  &  1/1 (100 \%)  \\
 13 & J1048-5832 &  1.2e-24   &  1.1e-23   &  0/3 (0 \%)  &  0/3 (0 \%)  &  2/3 (67 \%)   &  1/3 (33 \%)   &  2/3 (67 \%)   \\
 14 & J1052-5954 &  5.7e-26   &  1.1e-23   &  0/1 (0 \%)  &  0/1 (0 \%)  &  1/1 (100 \%)  &   0/1 (0 \%)   &  1/1 (100 \%)  \\
 15 & J1112-6103 &  1.9e-24   &  1.0e-23   &  0/2 (0 \%)  &  0/2 (0 \%)  &  2/2 (100 \%)  &  2/2 (100 \%)  &  2/2 (100 \%)  \\
 16 & J1119-6127 &  2.4e-25   &  3.9e-25   &  0/2 (0 \%)  &  0/2 (0 \%)  &   0/2 (0 \%)   &   0/2 (0 \%)   &   0/2 (0 \%)   \\
 17 & J1123-6259 &  5.4e-27   &  3.4e-24   &  0/1 (0 \%)  &  0/1 (0 \%)  &   0/1 (0 \%)   &   0/1 (0 \%)   &   0/1 (0 \%)   \\
 18 & J1141-6545 &  4.4e-27   &  4.3e-25   &  0/1 (0 \%)  &  0/1 (0 \%)  &   0/1 (0 \%)   &   0/1 (0 \%)   &   0/1 (0 \%)   \\
 19 & J1301-6305 &  1.4e-26   &  2.6e-24   &  0/1 (0 \%)  &  0/1 (0 \%)  &   0/1 (0 \%)   &   0/1 (0 \%)   &   0/1 (0 \%)   \\
 20 & J1302-6350 &  4.6e-26   &  8.0e-25   &  0/1 (0 \%)  &  0/1 (0 \%)  &   0/1 (0 \%)   &   0/1 (0 \%)   &   0/1 (0 \%)   \\
 21 & J1341-6220 &  2.8e-25   &  1.1e-24   &  0/4 (0 \%)  &  0/4 (0 \%)  &   0/4 (0 \%)   &   0/4 (0 \%)   &   0/4 (0 \%)   \\
 22 & J1406-6121 &  2.1e-25   &  5.9e-25   &  0/1 (0 \%)  &  0/1 (0 \%)  &   0/1 (0 \%)   &   0/1 (0 \%)   &   0/1 (0 \%)   \\
 23 & J1412-6145 &  1.9e-26   &  2.8e-24   &  0/1 (0 \%)  &  0/1 (0 \%)  &   0/1 (0 \%)   &   0/1 (0 \%)   &   0/1 (0 \%)   \\
 24 & J1420-6048 &  1.5e-24   &  8.6e-24   &  0/3 (0 \%)  &  0/3 (0 \%)  &  1/3 (33 \%)   &   0/3 (0 \%)   &  2/3 (67 \%)   \\
 25 & J1452-6036 &  8.7e-27   &  4.1e-25   &  0/1 (0 \%)  &  0/1 (0 \%)  &   0/1 (0 \%)   &   0/1 (0 \%)   &   0/1 (0 \%)   \\
 26 & J1524-5625 &  4.2e-26   &  1.5e-23   &  0/1 (0 \%)  &  0/1 (0 \%)  &  1/1 (100 \%)  &  1/1 (100 \%)  &  1/1 (100 \%)  \\
 27 & J1531-5610 &  6.5e-26   &  1.5e-23   &  0/1 (0 \%)  &  0/1 (0 \%)  &  1/1 (100 \%)  &  1/1 (100 \%)  &  1/1 (100 \%)  \\
 28 & J1617-5055 &  1.1e-24   &  7.6e-25   &  0/1 (0 \%)  &  0/1 (0 \%)  &   0/1 (0 \%)   &   0/1 (0 \%)   &   0/1 (0 \%)   \\
 29 & J1646-4346 &  3.8e-25   &  4.7e-24   &  0/1 (0 \%)  &  0/1 (0 \%)  &   0/1 (0 \%)   &   0/1 (0 \%)   &   0/1 (0 \%)   \\
 30 & J1650-4502 &  9.1e-26   &  5.4e-24   &  0/1 (0 \%)  &  0/1 (0 \%)  &   0/1 (0 \%)   &   0/1 (0 \%)   &   0/1 (0 \%)   \\
 31 & J1702-4128 &  1.8e-25   &  5.5e-24   &  0/1 (0 \%)  &  0/1 (0 \%)  &   0/1 (0 \%)   &   0/1 (0 \%)   &   0/1 (0 \%)   \\
 32 & J1702-4310 &  1.3e-24   &  4.9e-24   &  0/2 (0 \%)  &  0/2 (0 \%)  &   0/2 (0 \%)   &   0/2 (0 \%)   &   0/2 (0 \%)   \\
 33 & J1705-3950 &  9.2e-26   &  6.5e-24   &  0/1 (0 \%)  &  0/1 (0 \%)  &   0/1 (0 \%)   &   0/1 (0 \%)   &   0/1 (0 \%)   \\
 34 & J1708-4008 &  5.4e-25   &  1.2e-25   &  0/1 (0 \%)  &  0/1 (0 \%)  &   0/1 (0 \%)   &   0/1 (0 \%)   &   0/1 (0 \%)   \\
 35 & J1709-4429 &  7.6e-26   &  1.5e-23   &  0/5 (0 \%)  &  0/5 (0 \%)  &  5/5 (100 \%)  &  5/5 (100 \%)  &  5/5 (100 \%)  \\
 36 & J1722-3632 &  2.0e-27   &  2.4e-24   &  0/1 (0 \%)  &  0/1 (0 \%)  &   0/1 (0 \%)   &   0/1 (0 \%)   &   0/1 (0 \%)   \\
 37 & J1730-3350 &  1.9e-24   &  8.3e-24   &  0/3 (0 \%)  &  0/3 (0 \%)  &  2/3 (67 \%)   &   0/3 (0 \%)   &  3/3 (100 \%)  \\
 38 & J1731-4744 &  1.7e-24   &  1.4e-24   &  0/3 (0 \%)  &  0/3 (0 \%)  &   0/3 (0 \%)   &   0/3 (0 \%)   &   0/3 (0 \%)   \\
 39 & J1740-3015 &  5.1e-24   &  1.6e-23   &  0/6 (0 \%)  &  0/6 (0 \%)  &   0/6 (0 \%)   &   0/6 (0 \%)   &   0/6 (0 \%)   \\
 40 & J1757-2421 &  2.4e-26   &  8.7e-24   &  0/1 (0 \%)  &  0/1 (0 \%)  &   0/1 (0 \%)   &   0/1 (0 \%)   &   0/1 (0 \%)   \\
 41 & J1801-2304 &  4.4e-27   &  9.8e-25   &  0/1 (0 \%)  &  0/1 (0 \%)  &   0/1 (0 \%)   &   0/1 (0 \%)   &   0/1 (0 \%)   \\
 42 & J1801-2451 &  4.6e-24   &  9.4e-24   &  0/3 (0 \%)  &  0/3 (0 \%)  &  2/3 (67 \%)   &  1/3 (33 \%)   &  2/3 (67 \%)   \\
 43 & J1803-2137 &  1.7e-25   &  8.6e-24   &  0/4 (0 \%)  &  0/4 (0 \%)  &  1/4 (25 \%)   &   0/4 (0 \%)   &  4/4 (100 \%)  \\
 44 & J1809-1917 &  1.9e-26   &  1.1e-23   &  0/1 (0 \%)  &  0/1 (0 \%)  &  1/1 (100 \%)  &   0/1 (0 \%)   &  1/1 (100 \%)  \\
 45 & J1812-1718 &  4.8e-27   &  5.5e-26   &  0/1 (0 \%)  &  0/1 (0 \%)  &   0/1 (0 \%)   &   0/1 (0 \%)   &   0/1 (0 \%)   \\
 46 & J1826-1334 &  1.4e-25   &  1.2e-23   &  0/2 (0 \%)  &  0/2 (0 \%)  &  2/2 (100 \%)  &   0/2 (0 \%)   &  2/2 (100 \%)  \\
 47 & J1833-0827 &  9.1e-27   &  8.1e-24   &  0/1 (0 \%)  &  0/1 (0 \%)  &   0/1 (0 \%)   &   0/1 (0 \%)   &   0/1 (0 \%)   \\
 48 & J1837-0604 &  2.0e-25   &  5.8e-24   &  0/1 (0 \%)  &  0/1 (0 \%)  &   0/1 (0 \%)   &   0/1 (0 \%)   &   0/1 (0 \%)   \\
 49 & J1841-0425 &  1.6e-26   &  2.1e-24   &  0/1 (0 \%)  &  0/1 (0 \%)  &   0/1 (0 \%)   &   0/1 (0 \%)   &   0/1 (0 \%)   \\
 50 & J1841-0456 &  3.8e-25   &  5.3e-26   &  0/1 (0 \%)  &  0/1 (0 \%)  &   0/1 (0 \%)   &   0/1 (0 \%)   &   0/1 (0 \%)   \\
 51 & J1841-0524 &  1.1e-25   &  1.3e-24   &  0/2 (0 \%)  &  0/2 (0 \%)  &   0/2 (0 \%)   &   0/2 (0 \%)   &   0/2 (0 \%)   \\
 52 & J1846-0258 &     -      &     -      &  0/1 (0 \%)  &  0/1 (0 \%)  &   0/1 (0 \%)   &   0/1 (0 \%)   &   0/1 (0 \%)   \\
 53 & J1852-0635 &  4.2e-27   &  1.0e-24   &  0/1 (0 \%)  &  0/1 (0 \%)  &   0/1 (0 \%)   &   0/1 (0 \%)   &   0/1 (0 \%)   \\
 54 & J1853+0545 &  1.5e-27   &  1.0e-25   &  0/1 (0 \%)  &  0/1 (0 \%)  &   0/1 (0 \%)   &   0/1 (0 \%)   &   0/1 (0 \%)   \\
 55 & J1957+2831 &  7.0e-27   &  8.1e-26   &  0/1 (0 \%)  &  0/1 (0 \%)  &   0/1 (0 \%)   &   0/1 (0 \%)   &   0/1 (0 \%)   \\
 56 & J2301+5852 &  6.5e-25   &  1.9e-25   &  0/1 (0 \%)  &  0/1 (0 \%)  &   0/1 (0 \%)   &   0/1 (0 \%)   &   0/1 (0 \%)   \\
 57 & J2337+6151 &  7.1e-25   &  3.0e-23   &  0/1 (0 \%)  &  0/1 (0 \%)  &   0/1 (0 \%)   &  1/1 (100 \%)  &  1/1 (100 \%)  \\
\hline
\end{tabular}
\end{table*}

\begin{table*}
\centering
\caption{Per-pulsar detection percentage with equal energy fractions $\digamma_{\!\mathrm{CW}}=\digamma_{\!\mathrm{fmode}}=Q$. The maximum GW amplitudes are shown for each pulsar, maximizing across all glitches and samples for each glitch. The number of glitches with at least one sample above both CW and Burst thresholds are shown for the different detector configurations.}
\label{tab:CW_Q_fmode_Q}
\begin{tabular}{ccccrrrrr}
\hline
    &    PSR     &  Max $h_{0,\mathrm{CW}}$  &  Max $h_{\mathrm{rss}}\,[\mathrm{Hz}^{-1/2}]$  &      O4     &     O5     &   O5 + CE    &   O5 + ET    &   ET + CE    \\
\hline
 1  & J0146+6145 &  1.4e-23   &  1.3e-25   & 0/1 (0 \%)  & 0/1 (0 \%)  &  0/1 (0 \%)   &  0/1 (0 \%)   &  0/1 (0 \%)   \\
 2  & J0205+6449 &  4.7e-25   &  1.5e-23   & 0/1 (0 \%)  & 0/1 (0 \%)  &  0/1 (0 \%)   &  0/1 (0 \%)   & 1/1 (100 \%)  \\
 3  & J0358+5413 &  4.0e-26   &  1.0e-24   & 0/1 (0 \%)  & 0/1 (0 \%)  &  0/1 (0 \%)   &  0/1 (0 \%)   &  0/1 (0 \%)   \\
 4  & J0528+2200 &  4.7e-25   &  2.2e-26   & 0/2 (0 \%)  & 0/2 (0 \%)  &  0/2 (0 \%)   &  0/2 (0 \%)   &  0/2 (0 \%)   \\
 5  & J0534+2200 &  9.9e-24   &  5.5e-24   & 0/10 (0 \%) & 0/10 (0 \%) &  0/10 (0 \%)  &  0/10 (0 \%)  &  0/10 (0 \%)  \\
 6  & J0631+1036 &  3.3e-25   &  4.0e-25   & 0/2 (0 \%)  & 0/2 (0 \%)  &  0/2 (0 \%)   &  0/2 (0 \%)   &  0/2 (0 \%)   \\
 7  & J0729-1448 &  5.5e-24   &  8.3e-25   & 0/1 (0 \%)  & 0/1 (0 \%)  &  0/1 (0 \%)   &  0/1 (0 \%)   &  0/1 (0 \%)   \\
 8  & J0835-4510 &  2.9e-22   &  4.9e-23   & 0/19 (0 \%) & 1/19 (5 \%) & 17/19 (89 \%) & 17/19 (89 \%) & 17/19 (89 \%) \\
 9  & J0940-5428 &  3.9e-25   &  7.7e-24   & 0/1 (0 \%)  & 0/1 (0 \%)  & 1/1 (100 \%)  & 1/1 (100 \%)  & 1/1 (100 \%)  \\
 10 & J1015-5719 &  5.1e-26   &  9.4e-25   & 0/1 (0 \%)  & 0/1 (0 \%)  &  0/1 (0 \%)   &  0/1 (0 \%)   &  0/1 (0 \%)   \\
 11 & J1016-5857 &  6.5e-25   &  9.4e-25   & 0/2 (0 \%)  & 0/2 (0 \%)  &  0/2 (0 \%)   &  0/2 (0 \%)   &  0/2 (0 \%)   \\
 12 & J1028-5819 &  1.0e-25   &  2.9e-24   & 0/1 (0 \%)  & 0/1 (0 \%)  &  0/1 (0 \%)   &  0/1 (0 \%)   &  0/1 (0 \%)   \\
 13 & J1048-5832 &  1.7e-24   &  2.0e-24   & 0/3 (0 \%)  & 0/3 (0 \%)  &  0/3 (0 \%)   &  0/3 (0 \%)   &  0/3 (0 \%)   \\
 14 & J1052-5954 &  6.2e-26   &  7.9e-25   & 0/1 (0 \%)  & 0/1 (0 \%)  &  0/1 (0 \%)   &  0/1 (0 \%)   &  0/1 (0 \%)   \\
 15 & J1112-6103 &  2.2e-24   &  5.1e-24   & 0/2 (0 \%)  & 0/2 (0 \%)  &  0/2 (0 \%)   &  0/2 (0 \%)   &  0/2 (0 \%)   \\
 16 & J1119-6127 &  2.4e-25   &  7.9e-25   & 0/2 (0 \%)  & 0/2 (0 \%)  &  0/2 (0 \%)   &  0/2 (0 \%)   &  0/2 (0 \%)   \\
 17 & J1123-6259 &  5.7e-27   &  1.7e-25   & 0/1 (0 \%)  & 0/1 (0 \%)  &  0/1 (0 \%)   &  0/1 (0 \%)   &  0/1 (0 \%)   \\
 18 & J1141-6545 &  4.4e-27   &  3.2e-26   & 0/1 (0 \%)  & 0/1 (0 \%)  &  0/1 (0 \%)   &  0/1 (0 \%)   &  0/1 (0 \%)   \\
 19 & J1301-6305 &  1.5e-26   &  1.8e-25   & 0/1 (0 \%)  & 0/1 (0 \%)  &  0/1 (0 \%)   &  0/1 (0 \%)   &  0/1 (0 \%)   \\
 20 & J1302-6350 &  1.3e-26   &  6.0e-25   & 0/1 (0 \%)  & 0/1 (0 \%)  &  0/1 (0 \%)   &  0/1 (0 \%)   &  0/1 (0 \%)   \\
 21 & J1341-6220 &  2.1e-25   &  2.6e-25   & 0/4 (0 \%)  & 0/4 (0 \%)  &  0/4 (0 \%)   &  0/4 (0 \%)   &  0/4 (0 \%)   \\
 22 & J1406-6121 &  8.5e-26   &  1.5e-25   & 0/1 (0 \%)  & 0/1 (0 \%)  &  0/1 (0 \%)   &  0/1 (0 \%)   &  0/1 (0 \%)   \\
 23 & J1412-6145 &  1.8e-26   &  1.4e-25   & 0/1 (0 \%)  & 0/1 (0 \%)  &  0/1 (0 \%)   &  0/1 (0 \%)   &  0/1 (0 \%)   \\
 24 & J1420-6048 &  1.2e-24   &  1.0e-24   & 0/3 (0 \%)  & 0/3 (0 \%)  &  0/3 (0 \%)   &  0/3 (0 \%)   &  0/3 (0 \%)   \\
 25 & J1452-6036 &  1.1e-26   &  1.7e-25   & 0/1 (0 \%)  & 0/1 (0 \%)  &  0/1 (0 \%)   &  0/1 (0 \%)   &  0/1 (0 \%)   \\
 26 & J1524-5625 &  4.2e-26   &  1.2e-24   & 0/1 (0 \%)  & 0/1 (0 \%)  &  0/1 (0 \%)   &  0/1 (0 \%)   &  0/1 (0 \%)   \\
 27 & J1531-5610 &  7.1e-26   &  1.6e-24   & 0/1 (0 \%)  & 0/1 (0 \%)  &  0/1 (0 \%)   &  0/1 (0 \%)   &  0/1 (0 \%)   \\
 28 & J1617-5055 &  8.4e-25   &  4.0e-24   & 0/1 (0 \%)  & 0/1 (0 \%)  &  0/1 (0 \%)   &  0/1 (0 \%)   &  0/1 (0 \%)   \\
 29 & J1646-4346 &  2.9e-25   &  4.1e-25   & 0/1 (0 \%)  & 0/1 (0 \%)  &  0/1 (0 \%)   &  0/1 (0 \%)   &  0/1 (0 \%)   \\
 30 & J1650-4502 &  1.2e-25   &  4.3e-25   & 0/1 (0 \%)  & 0/1 (0 \%)  &  0/1 (0 \%)   &  0/1 (0 \%)   &  0/1 (0 \%)   \\
 31 & J1702-4128 &  1.6e-25   &  3.7e-25   & 0/1 (0 \%)  & 0/1 (0 \%)  &  0/1 (0 \%)   &  0/1 (0 \%)   &  0/1 (0 \%)   \\
 32 & J1702-4310 &  4.1e-25   &  8.6e-25   & 0/2 (0 \%)  & 0/2 (0 \%)  &  0/2 (0 \%)   &  0/2 (0 \%)   &  0/2 (0 \%)   \\
 33 & J1705-3950 &  7.6e-26   &  4.8e-25   & 0/1 (0 \%)  & 0/1 (0 \%)  &  0/1 (0 \%)   &  0/1 (0 \%)   &  0/1 (0 \%)   \\
 34 & J1708-4008 &  7.5e-25   &  1.4e-25   & 0/1 (0 \%)  & 0/1 (0 \%)  &  0/1 (0 \%)   &  0/1 (0 \%)   &  0/1 (0 \%)   \\
 35 & J1709-4429 &  1.0e-25   &  2.5e-24   & 0/5 (0 \%)  & 0/5 (0 \%)  &  0/5 (0 \%)   &  0/5 (0 \%)   &  0/5 (0 \%)   \\
 36 & J1722-3632 &  1.8e-27   &  2.1e-26   & 0/1 (0 \%)  & 0/1 (0 \%)  &  0/1 (0 \%)   &  0/1 (0 \%)   &  0/1 (0 \%)   \\
 37 & J1730-3350 &  6.3e-25   &  9.6e-25   & 0/3 (0 \%)  & 0/3 (0 \%)  &  0/3 (0 \%)   &  0/3 (0 \%)   &  0/3 (0 \%)   \\
 38 & J1731-4744 &  5.9e-24   &  9.2e-25   & 0/3 (0 \%)  & 0/3 (0 \%)  &  0/3 (0 \%)   &  0/3 (0 \%)   &  0/3 (0 \%)   \\
 39 & J1740-3015 &  2.5e-24   &  2.2e-24   & 0/6 (0 \%)  & 0/6 (0 \%)  &  0/6 (0 \%)   &  0/6 (0 \%)   &  0/6 (0 \%)   \\
 40 & J1757-2421 &  3.0e-26   &  4.3e-25   & 0/1 (0 \%)  & 0/1 (0 \%)  &  0/1 (0 \%)   &  0/1 (0 \%)   &  0/1 (0 \%)   \\
 41 & J1801-2304 &  4.5e-27   &  1.1e-25   & 0/1 (0 \%)  & 0/1 (0 \%)  &  0/1 (0 \%)   &  0/1 (0 \%)   &  0/1 (0 \%)   \\
 42 & J1801-2451 &  4.4e-25   &  1.6e-24   & 0/3 (0 \%)  & 0/3 (0 \%)  &  0/3 (0 \%)   &  0/3 (0 \%)   &  0/3 (0 \%)   \\
 43 & J1803-2137 &  1.6e-25   &  1.2e-24   & 0/4 (0 \%)  & 0/4 (0 \%)  &  0/4 (0 \%)   &  0/4 (0 \%)   &  0/4 (0 \%)   \\
 44 & J1809-1917 &  1.9e-26   &  8.4e-25   & 0/1 (0 \%)  & 0/1 (0 \%)  &  0/1 (0 \%)   &  0/1 (0 \%)   &  0/1 (0 \%)   \\
 45 & J1812-1718 &  5.0e-27   &  3.5e-26   & 0/1 (0 \%)  & 0/1 (0 \%)  &  0/1 (0 \%)   &  0/1 (0 \%)   &  0/1 (0 \%)   \\
 46 & J1826-1334 &  1.0e-24   &  9.8e-25   & 0/2 (0 \%)  & 0/2 (0 \%)  &  0/2 (0 \%)   &  0/2 (0 \%)   &  0/2 (0 \%)   \\
 47 & J1833-0827 &  6.5e-27   &  3.1e-25   & 0/1 (0 \%)  & 0/1 (0 \%)  &  0/1 (0 \%)   &  0/1 (0 \%)   &  0/1 (0 \%)   \\
 48 & J1837-0604 &  2.3e-25   &  1.9e-24   & 0/1 (0 \%)  & 0/1 (0 \%)  &  0/1 (0 \%)   &  0/1 (0 \%)   &  0/1 (0 \%)   \\
 49 & J1841-0425 &  1.3e-26   &  1.3e-25   & 0/1 (0 \%)  & 0/1 (0 \%)  &  0/1 (0 \%)   &  0/1 (0 \%)   &  0/1 (0 \%)   \\
 50 & J1841-0456 &  3.7e-25   &  6.7e-26   & 0/1 (0 \%)  & 0/1 (0 \%)  &  0/1 (0 \%)   &  0/1 (0 \%)   &  0/1 (0 \%)   \\
 51 & J1841-0524 &  7.4e-26   &  2.6e-25   & 0/2 (0 \%)  & 0/2 (0 \%)  &  0/2 (0 \%)   &  0/2 (0 \%)   &  0/2 (0 \%)   \\
 52 & J1846-0258 &  3.4e-25   &  3.4e-24   & 0/1 (0 \%)  & 0/1 (0 \%)  &  0/1 (0 \%)   &  0/1 (0 \%)   &  0/1 (0 \%)   \\
 53 & J1852-0635 &  4.2e-27   &  4.8e-26   & 0/1 (0 \%)  & 0/1 (0 \%)  &  0/1 (0 \%)   &  0/1 (0 \%)   &  0/1 (0 \%)   \\
 54 & J1853+0545 &  1.5e-27   &  5.4e-26   & 0/1 (0 \%)  & 0/1 (0 \%)  &  0/1 (0 \%)   &  0/1 (0 \%)   &  0/1 (0 \%)   \\
 55 & J1957+2831 &  7.2e-27   &  7.7e-26   & 0/1 (0 \%)  & 0/1 (0 \%)  &  0/1 (0 \%)   &  0/1 (0 \%)   &  0/1 (0 \%)   \\
 56 & J2301+5852 &  7.1e-25   &  1.3e-25   & 0/1 (0 \%)  & 0/1 (0 \%)  &  0/1 (0 \%)   &  0/1 (0 \%)   &  0/1 (0 \%)   \\
 57 & J2337+6151 &  7.6e-25   &  2.3e-24   & 0/1 (0 \%)  & 0/1 (0 \%)  &  0/1 (0 \%)   &  0/1 (0 \%)   &  0/1 (0 \%)   \\
\hline
\end{tabular}
\end{table*}

\bibliography{references}

@article{O4_vela_glitch_search,
	title = {Constraints on {Gravitational} {Waves} from the 2024 {Vela} {Pulsar} {Glitch}},
	volume = {1005},
	issn = {0004-637X, 1538-4357},
	url = {https://iopscience.iop.org/article/10.3847/1538-4357/ae63bc},
	doi = {10.3847/1538-4357/ae63bc},
	number = {1},
	urldate = {2026-07-07},
	journal = {The Astrophysical Journal},
	author = {Abac, A. G. and others},
	collaboration = "LIGO Scientific Collaboration, Virgo Collaboration, and KAGRA Collaboration",
	month = {07},
	year = {2026},
	pages = {12},
}

@misc{vela_glitch_zenodo,
	title = {Data release for paper "Constraints on gravitational waves from the 2024 Vela pulsar glitch"},
	author = "{LIGO Scientific Collaboration} and {Virgo Collaboration} and {KAGRA Collaboration}",
	howpublished   = "Zenodo",
	publisher    = {Zenodo},
	doi = {https://doi.org/10.5281/zenodo.17735648},
	year = {2025},
	month = {12},
	url = {https://zenodo.org/records/17735649},
	note = {\url{https://zenodo.org/records/17735649}},
}

@article{ball_methods_2025,
	author = {{Ball}, Matthew and {Frey}, Raymond},
        title = "{Methods for estimating neutron star parameters using multiple mechanisms for gravitational wave emission associated with pulsar glitches}",
      journal = {Monthly Notices of the Royal Astronomical Society},
         year = 2026,
        month = {01},
       volume = {545},
       number = {3},
          eid = {staf2176},
        pages = {staf2176},
          doi = {10.1093/mnras/staf2176},
archivePrefix = {arXiv},
       eprint = {2508.15965},
 primaryClass = {astro-ph.HE},
       adsurl = {https://ui.adsabs.harvard.edu/abs/2026MNRAS.545f2176B}
}

@article{moragues_prospects_2023,
	title = {Prospects for detecting transient quasi-monochromatic gravitational waves from glitching pulsars with current and future detectors},
	volume = {519},
	copyright = {https://academic.oup.com/journals/pages/open\_access/funder\_policies/chorus/standard\_publication\_model},
	issn = {0035-8711, 1365-2966},
	url = {https://academic.oup.com/mnras/article/519/4/5161/6912269},
	doi = {10.1093/mnras/stac3665},
	number = {4},
	urldate = {2026-03-04},
	journal = {Monthly Notices of the Royal Astronomical Society},
	author = {Moragues, Joan and Modafferi, Luana M and Tenorio, Rodrigo and Keitel, David},
	month = {01},
	year = {2023},
	pages = {5161--5176},
}

@misc{prix_2011,
  author = {Prix, Reinhard},
  howpublished = {LIGO Document Control Center},
  title = {The F-statistic and its implementation in ComputeFstatistic v2},
  year = {2018},
  url = {https://https://dcc.ligo.org/T0900149/public},
  number = {LIGO-T0900149},
  institution = {LIGO Laboratory}
}

@article{Mirasola_2024lcq,
    author = "Mirasola, Lorenzo and Tenorio, Rodrigo",
    title = "{Toward a computationally efficient follow-up pipeline for blind continuous gravitational-wave searches}",
    eprint = "2405.18934",
    archivePrefix = "arXiv",
    primaryClass = "gr-qc",
    reportNumber = "LIGO-P2400221",
    doi = "10.1103/PhysRevD.110.124049",
    journal = "Phys. Rev. D",
    volume = "110",
    number = "12",
    pages = "124049",
    year = "2024"
}

@article{KAGRA:2013rdx,
    author = "Abbott, B. P. and others",
    collaboration = "LIGO Scientific Collaboration, Virgo Collaboration, and KAGRA Collaboration",
    title = "{Prospects for observing and localizing gravitational-wave transients with Advanced LIGO, Advanced Virgo and KAGRA}",
    eprint = "1304.0670",
    archivePrefix = "arXiv",
    primaryClass = "gr-qc",
    reportNumber = "LIGO-P1200087, VIR-0288A-12, JGW-P1808427",
    doi = "10.1007/s41114-020-00026-9",
    journal = "Living Rev. Rel.",
    volume = "23",
	number = "1",
    pages = "3",
    year = "2020"
}

@article{LIGOScientific:2014pky,
    author = "Aasi, J. and others",
    collaboration = "LIGO Scientific Collaboration",
    title = "{Advanced LIGO}",
    eprint = "1411.4547",
    archivePrefix = "arXiv",
    primaryClass = "gr-qc",
    doi = "10.1088/0264-9381/32/7/074001",
    journal = "Class. Quant. Grav.",
    volume = "32",
    pages = "074001",
    year = "2015"
}

@article{VIRGO:2014yos,
    author = "Acernese, F. and others",
    collaboration = "Virgo",
    title = "{Advanced Virgo: a second-generation interferometric gravitational wave detector}",
    eprint = "1408.3978",
    archivePrefix = "arXiv",
    primaryClass = "gr-qc",
    doi = "10.1088/0264-9381/32/2/024001",
    journal = "Class. Quant. Grav.",
    volume = "32",
    number = "2",
    pages = "024001",
    year = "2015"
}

@article{10.1093/ptep/ptaa125,
    author = {Akutsu, T and others},
    title = {Overview of KAGRA: Detector design and construction history},
    journal = {Progress of Theoretical and Experimental Physics},
    volume = {2021},
    number = {5},
    pages = {05A101},
    year = {2021},
    month = {05},
    issn = {2050-3911},
    doi = {10.1093/ptep/ptaa125},
    url = {https://doi.org/10.1093/ptep/ptaa125},
    eprint = {https://academic.oup.com/ptep/article-pdf/2021/5/05A101/37974994/ptaa125.pdf},
}

@book{lorimer_handbook_2004,
	address = {Cambridge},
	series = {Cambridge {Observing} {Handbooks} for {Research} {Astronomers}},
	title = {Handbook of {Pulsar} {Astronomy}},
	volume = {4},
	isbn = {978-0-521-82823-9},
	urldate = {2026-02-12},
	publisher = {Cambridge University Press},
	author = {Lorimer, D. R. and Kramer, M.},
	month = {12},
	year = {2004},
}

@article{fuentes_glitch_2017,
	title = {The glitch activity of neutron stars},
	volume = {608},
	issn = {0004-6361, 1432-0746},
	url = {https://www.aanda.org/10.1051/0004-6361/201731519},
	doi = {10.1051/0004-6361/201731519},
	urldate = {2026-02-12},
	journal = {Astronomy \& Astrophysics},
	author = {Fuentes, J. R. and Espinoza, C. M. and Reisenegger, A. and Shaw, B. and Stappers, B. W. and Lyne, A. G.},
	month = {12},
	year = {2017},
	pages = {A131},
}

@article{antonopoulou_pulsar_2022,
	title = {Pulsar glitches: observations and physical interpretation},
	volume = {85},
	issn = {0034-4885, 1361-6633},
	shorttitle = {Pulsar glitches},
	url = {https://iopscience.iop.org/article/10.1088/1361-6633/ac9ced},
	doi = {10.1088/1361-6633/ac9ced},
	number = {12},
	urldate = {2026-02-12},
	journal = {Reports on Progress in Physics},
	author = {Antonopoulou, Danai and Haskell, Brynmor and Espinoza, Cristóbal M},
	month = {12},
	year = {2022},
	pages = {126901},
}

@article{haskell_models_2015,
	title = {Models of pulsar glitches},
	volume = {24},
	issn = {0218-2718, 1793-6594},
	url = {https://www.worldscientific.com/doi/abs/10.1142/S0218271815300086},
	doi = {10.1142/S0218271815300086},
	number = {03},
	urldate = {2026-02-12},
	journal = {International Journal of Modern Physics D},
	author = {Haskell, Brynmor and Melatos, Andrew},
	month = {03},
	year = {2015},
	pages = {1530008},
}

@article{zhou_pulsar_2022,
	title = {Pulsar {Glitches}: {A} {Review}},
	volume = {8},
	issn = {2218-1997},
	shorttitle = {Pulsar {Glitches}},
	url = {https://www.mdpi.com/2218-1997/8/12/641},
	doi = {10.3390/universe8120641},
	number = {12},
	urldate = {2026-02-12},
	journal = {Universe},
	author = {Zhou, Shiqi and Gügercinoğlu, Erbil and Yuan, Jianping and Ge, Mingyu and Yu, Cong},
	month = {12},
	year = {2022},
	pages = {641},
}

@article{link_pulsar_1992,
	title = {Pulsar glitches as probes of neutron star interiors},
	volume = {359},
	copyright = {http://www.springer.com/tdm},
	issn = {0028-0836, 1476-4687},
	url = {https://www.nature.com/articles/359616a0},
	doi = {10.1038/359616a0},
	number = {6396},
	urldate = {2026-02-12},
	journal = {Nature},
	author = {Link, Bennett and Epstein, Richard I. and Van Riper, Kenneth A.},
	month = {10},
	year = {1992},
	pages = {616--618},
}

@incollection{cheng_probing_2000,
	address = {Dordrecht},
	title = {Probing the {Neutron} {Star} {Interior} with {Glitches}},
	volume = {254},
	isbn = {978-94-010-3791-4 978-94-010-0878-5},
	url = {http://link.springer.com/10.1007/978-94-010-0878-5_14},
	urldate = {2026-02-12},
	booktitle = {Stellar {Astrophysics}},
	publisher = {Springer Netherlands},
	author = {Link, Bennett and Epstein, Richard I. and Lattimer, James M.},
	editor = {Cheng, K. S. and Chau, H. F. and Chan, K. L. and Leung, K. C.},
	year = {2000},
	doi = {10.1007/978-94-010-0878-5_14},
	note = {Series Title: Astrophysics and Space Science Library},
	pages = {117--126},
}

@article{haskell_probing_2017,
	title = {Probing neutron star interiors with pulsar glitches},
	volume = {13},
	copyright = {https://www.cambridge.org/core/terms},
	issn = {1743-9213, 1743-9221},
	url = {https://www.cambridge.org/core/product/identifier/S1743921317010663/type/journal_article},
	doi = {10.1017/S1743921317010663},
	number = {S337},
	urldate = {2026-02-12},
	journal = {Proceedings of the International Astronomical Union},
	author = {Haskell, Brynmor},
	month = {09},
	year = {2017},
	pages = {203--208},
}

@article{prix_search_2011,
	title = {Search method for long-duration gravitational-wave transients from neutron stars},
	volume = {84},
	copyright = {http://link.aps.org/licenses/aps-default-license},
	issn = {1550-7998, 1550-2368},
	url = {https://link.aps.org/doi/10.1103/PhysRevD.84.023007},
	doi = {10.1103/PhysRevD.84.023007},
	number = {2},
	urldate = {2026-02-12},
	journal = {Physical Review D},
	author = {Prix, R. and Giampanis, S. and Messenger, C.},
	month = {07},
	year = {2011},
	pages = {023007},
}

@article{keitel_first_2019,
	title = {First search for long-duration transient gravitational waves after glitches in the {Vela} and {Crab} pulsars},
	volume = {100},
	issn = {2470-0010, 2470-0029},
	url = {https://link.aps.org/doi/10.1103/PhysRevD.100.064058},
	doi = {10.1103/PhysRevD.100.064058},
	number = {6},
	urldate = {2026-02-12},
	journal = {Physical Review D},
	author = {Keitel, David and Woan, Graham and Pitkin, Matthew and Schumacher, Courtney and Pearlstone, Brynley and Riles, Keith and Lyne, Andrew G. and Palfreyman, Jim and Stappers, Benjamin and Weltevrede, Patrick},
	month = {09},
	year = {2019},
	pages = {064058},
}

@article{abbott_all-sky_2021,
	title = {All-sky search for short gravitational-wave bursts in the third {Advanced} {LIGO} and {Advanced} {Virgo} run},
	collaboration = "LIGO Scientific Collaboration, Virgo Collaboration, and KAGRA Collaboration",
	volume = {104},
	issn = {2470-0010, 2470-0029},
	url = {https://link.aps.org/doi/10.1103/PhysRevD.104.122004},
	doi = {10.1103/PhysRevD.104.122004},
	number = {12},
	urldate = {2026-02-12},
	journal = {Physical Review D},
	author = {Abbott, R. and others},
	month = {12},
	year = {2021},
	pages = {122004},
}

@article{lopez_prospects_2022,
	title = {Prospects for detecting and localizing short-duration transient gravitational waves from glitching neutron stars without electromagnetic counterparts},
	volume = {106},
	issn = {2470-0010, 2470-0029},
	url = {https://link.aps.org/doi/10.1103/PhysRevD.106.103037},
	doi = {10.1103/PhysRevD.106.103037},
	number = {10},
	urldate = {2026-02-12},
	journal = {Physical Review D},
	author = {Lopez, Dixeena and Tiwari, Shubhanshu and Drago, Marco and Keitel, David and Lazzaro, Claudia and Prodi, Giovanni Andrea},
	month = {11},
	year = {2022},
	pages = {103037},
}

@article{lopez_search_2024,
	title = {Search for {Short}-{Duration} {Transient} {Gravitational} {Waves} {Emitted} by {Neutron} {Star} {Glitches}},
	volume = {536},
	issn = {0003-3804, 1521-3889},
	url = {https://onlinelibrary.wiley.com/doi/10.1002/andp.202200142},
	doi = {10.1002/andp.202200142},
	number = {2},
	urldate = {2026-02-12},
	journal = {Annalen der Physik},
	author = {Lopez, Dixeena and {the LIGO, Virgo, and KAGRA collaborations}},
	month = {02},
	year = {2024},
	pages = {2200142},
}

@article{abbott_all-sky_2017,
	title = {All-sky search for short gravitational-wave bursts in the first {Advanced} {LIGO} run},
	volume = {95},
	copyright = {https://link.aps.org/licenses/aps-default-license},
	issn = {2470-0010, 2470-0029},
	url = {https://link.aps.org/doi/10.1103/PhysRevD.95.042003},
	doi = {10.1103/PhysRevD.95.042003},
	number = {4},
	urldate = {2026-02-12},
	journal = {Physical Review D},
	author = {Abbott, B.P. and others},
	collaboration = "LIGO Scientific Collaboration and Virgo Collaboration",
	month = {02},
	year = {2017},
	pages = {042003},
}

@article{abbott_all-sky_2019,
	title = {All-sky search for short gravitational-wave bursts in the second {Advanced} {LIGO} and {Advanced} {Virgo} run},
	volume = {100},
	issn = {2470-0010, 2470-0029},
	url = {https://link.aps.org/doi/10.1103/PhysRevD.100.024017},
	doi = {10.1103/PhysRevD.100.024017},
	number = {2},
	urldate = {2026-02-12},
	journal = {Physical Review D},
	author = {Abbott, B.P. and others},
	collaboration = "LIGO Scientific Collaboration and Virgo Collaboration",
	month = {07},
	year = {2019},
	pages = {024017},
}

@article{abac_all-sky_2025,
	title = {All-sky search for short gravitational-wave bursts in the first part of the fourth {LIGO}-{Virgo}-{KAGRA} observing run},
	volume = {112},
	issn = {2470-0010, 2470-0029},
	url = {https://link.aps.org/doi/10.1103/wjdz-jdby},
	doi = {10.1103/wjdz-jdby},
	number = {10},
	urldate = {2026-02-12},
	journal = {Physical Review D},
	author = {Abac, A.G. and others},
	collaboration = "LIGO Scientific Collaboration, Virgo Collaboration, and KAGRA Collaboration",
	month = {11},
	year = {2025},
	pages = {102005},
}

@article{abadie_search_2011,
	title = {Search for gravitational waves associated with the {August} 2006 timing glitch of the {Vela} pulsar},
	volume = {83},
	copyright = {http://link.aps.org/licenses/aps-default-license},
	issn = {1550-7998, 1550-2368},
	url = {https://link.aps.org/doi/10.1103/PhysRevD.83.042001},
	doi = {10.1103/PhysRevD.83.042001},
	number = {4},
	urldate = {2026-02-12},
	journal = {Physical Review D},
	author = {Abadie, J. and others},
	collaboration = {LIGO Scientific Collaboration},
	month = {02},
	year = {2011},
	pages = {042001},
}

@article{van_eysden_gravitational_2008,
	title = {Gravitational radiation from pulsar glitches},
	volume = {25},
	issn = {0264-9381, 1361-6382},
	url = {https://iopscience.iop.org/article/10.1088/0264-9381/25/22/225020},
	doi = {10.1088/0264-9381/25/22/225020},
	number = {22},
	urldate = {2026-02-12},
	journal = {Classical and Quantum Gravity},
	author = {Van Eysden, C A and Melatos, A},
	month = {11},
	year = {2008},
	pages = {225020},
}

@article{bennett_continuous-wave_2010,
	title = {Continuous-wave gravitational radiation from pulsar glitch recovery: {Gravitational} radiation from glitch recovery},
	volume = {409},
	issn = {00358711},
	shorttitle = {Continuous-wave gravitational radiation from pulsar glitch recovery},
	url = {https://academic.oup.com/mnras/article-lookup/doi/10.1111/j.1365-2966.2010.17416.x},
	doi = {10.1111/j.1365-2966.2010.17416.x},
	number = {4},
	urldate = {2026-02-12},
	journal = {Monthly Notices of the Royal Astronomical Society},
	author = {Bennett, M. F. and Van Eysden, C. A. and Melatos, A.},
	month = {12},
	year = {2010},
	pages = {1705--1718},
}

@article{melatos_persistent_2015,
	title = {Persistent Gravitational Radiation from Glitching Pulsars},
	volume = {807},
	copyright = {http://iopscience.iop.org/info/page/text-and-data-mining},
	issn = {1538-4357},
	url = {https://iopscience.iop.org/article/10.1088/0004-637X/807/2/132},
	doi = {10.1088/0004-637X/807/2/132},
	number = {2},
	urldate = {2026-02-12},
	journal = {The Astrophysical Journal},
	author = {Melatos, A. and Douglass, J. A. and Simula, T. P.},
	month = {07},
	year = {2015},
	pages = {132},
}

@article{singh_gravitational_2017,
	title = {Gravitational wave transient signal emission via {Ekman} pumping in neutron stars during post-glitch relaxation phase},
	volume = {95},
	copyright = {http://link.aps.org/licenses/aps-default-license},
	issn = {2470-0010, 2470-0029},
	url = {https://link.aps.org/doi/10.1103/PhysRevD.95.024022},
	doi = {10.1103/PhysRevD.95.024022},
	number = {2},
	urldate = {2026-02-12},
	journal = {Physical Review D},
	author = {Singh, Avneet},
	month = {01},
	year = {2017},
	pages = {024022},
}

@article{yim_transient_2020,
	title = {Transient gravitational waves from pulsar post-glitch recoveries},
	volume = {498},
	copyright = {https://academic.oup.com/journals/pages/open\_access/funder\_policies/chorus/standard\_publication\_model},
	issn = {0035-8711, 1365-2966},
	url = {https://academic.oup.com/mnras/article/498/3/3138/5895343},
	doi = {10.1093/mnras/staa2534},
	number = {3},
	urldate = {2026-02-12},
	journal = {Monthly Notices of the Royal Astronomical Society},
	author = {Yim, Garvin and Jones, D I},
	month = {09},
	year = {2020},
	pages = {3138--3152},
}

@article{10.1093/mnras/stae1659,
    author = {Yim, Garvin and Shao, Lijing and Xu, Renxin},
    title = {High-priority targets for transient gravitational waves from glitching pulsars},
    journal = {Monthly Notices of the Royal Astronomical Society},
    volume = {532},
    number = {4},
    pages = {3893-3913},
    year = {2024},
    month = {08},
    issn = {0035-8711},
    doi = {10.1093/mnras/stae1659},
    url = {https://doi.org/10.1093/mnras/stae1659},
    eprint = {https://academic.oup.com/mnras/article-pdf/532/4/3893/58691262/stae1659.pdf},
}

@article{yagi_approximate_2017,
	title = {Approximate universal relations for neutron stars and quark stars},
	volume = {681},
	issn = {03701573},
	url = {https://linkinghub.elsevier.com/retrieve/pii/S0370157317300492},
	doi = {10.1016/j.physrep.2017.03.002},
	urldate = {2026-02-13},
	journal = {Physics Reports},
	author = {Yagi, Kent and Yunes, Nicolás},
	month = {04},
	year = {2017},
	pages = {1--72},
}

@article{yang_exploring_2022,
	title = {Exploring the universal relations with the correlation analysis of neutron star properties},
	volume = {105},
	issn = {2470-0010, 2470-0029},
	url = {https://link.aps.org/doi/10.1103/PhysRevD.105.063023},
	doi = {10.1103/PhysRevD.105.063023},
	number = {6},
	urldate = {2026-04-22},
	journal = {Physical Review D},
	author = {Yang, Shen and Wen, Dehua and Wang, Jue and Zhang, Jing},
	month = {03},
	year = {2022},
	pages = {063023},
}

@article{echeverria_gravitational-wave_1989,
	title = {Gravitational-wave measurements of the mass and angular momentum of a black hole},
	volume = {40},
	copyright = {http://link.aps.org/licenses/aps-default-license},
	issn = {0556-2821},
	url = {https://link.aps.org/doi/10.1103/PhysRevD.40.3194},
	doi = {10.1103/PhysRevD.40.3194},
	number = {10},
	urldate = {2026-02-13},
	journal = {Physical Review D},
	author = {Echeverria, Fernando},
	month = {11},
	year = {1989},
	pages = {3194--3203},
}

@article{finn_detection_1992,
	title = {Detection, measurement, and gravitational radiation},
	volume = {46},
	copyright = {http://link.aps.org/licenses/aps-default-license},
	issn = {0556-2821},
	url = {https://link.aps.org/doi/10.1103/PhysRevD.46.5236},
	doi = {10.1103/PhysRevD.46.5236},
	number = {12},
	urldate = {2026-02-13},
	journal = {Physical Review D},
	author = {Finn, Lee S.},
	month = {12},
	year = {1992},
	pages = {5236--5249},
}

@article{pradhan_general_2022,
	title = {General relativistic treatment of f -mode oscillations of hyperonic stars},
	volume = {106},
	issn = {2469-9985, 2469-9993},
	url = {https://link.aps.org/doi/10.1103/PhysRevC.106.015805},
	doi = {10.1103/PhysRevC.106.015805},
	number = {1},
	urldate = {2026-02-13},
	journal = {Physical Review C},
	author = {Pradhan, Bikram Keshari and Chatterjee, Debarati and Lanoye, Michael and Jaikumar, Prashanth},
	month = {07},
	year = {2022},
	pages = {015805},
}

@article{kruger_fast_2020,
	title = {Fast {Rotating} {Relativistic} {Stars}: {Spectra} and {Stability} without {Approximation}},
	volume = {125},
	issn = {0031-9007, 1079-7114},
	shorttitle = {Fast {Rotating} {Relativistic} {Stars}},
	url = {https://link.aps.org/doi/10.1103/PhysRevLett.125.111106},
	doi = {10.1103/PhysRevLett.125.111106},
	number = {11},
	urldate = {2026-02-13},
	journal = {Physical Review Letters},
	author = {Krüger, Christian J. and Kokkotas, Kostas D.},
	month = {09},
	year = {2020},
	pages = {111106},
}

@misc{ATNF_catalog,
	author = {ATNF},
	title = {The ATNF Pulsar Catalogue Glitch Database},
	year = {2026},
	note = {\url{https://www.atnf.csiro.au/research/pulsar/psrcat/glitchTbl.html} [Accessed: 2026-01-14]}
}

@article{manchester_australia_2005,
	title = {The {Australia} {Telescope} {National} {Facility} {Pulsar} {Catalogue}},
	volume = {129},
	issn = {0004-6256, 1538-3881},
	url = {https://iopscience.iop.org/article/10.1086/428488},
	doi = {10.1086/428488},
	number = {4},
	urldate = {2026-02-20},
	journal = {The Astronomical Journal},
	author = {Manchester, R. N. and Hobbs, G. B. and Teoh, A. and Hobbs, M.},
	month = {04},
	year = {2005},
	pages = {1993--2006},
}

@article{ng_fitting_2008,
	title = {Fitting {Pulsar} {Wind} {Tori}. {II}. {Error} {Analysis} and {Applications}},
	volume = {673},
	issn = {0004-637X, 1538-4357},
	url = {https://iopscience.iop.org/article/10.1086/523935},
	doi = {10.1086/523935},
	number = {1},
	urldate = {2026-02-16},
	journal = {The Astrophysical Journal},
	author = {Ng, C.‐Y. and Romani, Roger W.},
	month = {01},
	year = {2008},
	pages = {411--417},
}

@article{lower_utmost_2020,
	title = {The {UTMOST} pulsar timing programme – {II}. {Timing} noise across the pulsar population},
	volume = {494},
	copyright = {https://academic.oup.com/journals/pages/open\_access/funder\_policies/chorus/standard\_publication\_model},
	issn = {0035-8711, 1365-2966},
	url = {https://academic.oup.com/mnras/article/494/1/228/5780241},
	doi = {10.1093/mnras/staa615},
	number = {1},
	urldate = {2026-02-20},
	journal = {Monthly Notices of the Royal Astronomical Society},
	author = {Lower, M E and Bailes, M and Shannon, R M and Johnston, S and Flynn, C and Osłowski, S and Gupta, V and Farah, W and Bateman, T and Green, A J and Hunstead, R and Jameson, A and Jankowski, F and Parthasarathy, A and Price, D C and Sutherland, A and Temby, D and Venkatraman Krishnan, V},
	month = {05},
	year = {2020},
	pages = {228--245},
}

@article{zubieta_first_2023,
	title = {First results of the glitching pulsar monitoring programme at the {Argentine} {Institute} of {Radioastronomy}},
	volume = {521},
	copyright = {https://academic.oup.com/journals/pages/open\_access/funder\_policies/chorus/standard\_publication\_model},
	issn = {0035-8711, 1365-2966},
	url = {https://academic.oup.com/mnras/article/521/3/4504/7074572},
	doi = {10.1093/mnras/stad723},
	number = {3},
	urldate = {2026-02-20},
	journal = {Monthly Notices of the Royal Astronomical Society},
	author = {Zubieta, Ezequiel and Missel, Ryan and Sosa Fiscella, Valentina and Lousto, Carlos O and del Palacio, Santiago and López Armengol, Federico G and García, Federico and Combi, Jorge A and Wang, Linwei and Combi, Luciano and Gancio, Guillermo and Negrelli, Carolina and Gutiérrez, Eduardo M},
	month = {03},
	year = {2023},
	pages = {4504--4521},
}

@article{atel16608,
	adsurl = {https://ui.adsabs.harvard.edu/abs/2024ATel16608....1Z},
 	author = {{Zubieta}, E. and {Furlan}, S.~B. Araujo and {Palacio}, S. del and {Garcia}, F. and {Gancio}, G. and {Lousto}, C.~O. and {Combi}, J.~A. and {PuMA Collaboration}},
	journal = {The Astronomer's Telegram},
	month = {May},
	pages = {1},
	title = {{Detection of a new giant glitch in the Vela Pulsar observed from the Argentine Institute of Radio astronomy}},
	volume = {16608},
	year = {2024},
	url = {https://www.astronomerstelegram.org/?read=16608}
}

@article{zubieta_study_2025,
	title = {Study of the 2024 major {Vela} glitch at the {Argentine} {Institute} of {Radioastronomy}},
	volume = {698},
	copyright = {https://creativecommons.org/licenses/by/4.0},
	issn = {0004-6361, 1432-0746},
	url = {https://www.aanda.org/10.1051/0004-6361/202554098},
	doi = {10.1051/0004-6361/202554098},
	urldate = {2026-02-20},
	journal = {Astronomy \& Astrophysics},
	author = {Zubieta, Ezequiel and Missel, Ryan and Araujo Furlan, Susana B. and Lousto, Carlos O. and García, Federico and Del Palacio, Santiago and Gancio, Guillermo and Combi, Jorge A. and Wang, Linwei},
	month = {06},
	year = {2025},
	pages = {A72},
}

@misc{lalsuite_2018,
    author         = "{LIGO Scientific Collaboration}
                      and {Virgo Collaboration}
                      and {KAGRA Collaboration}",
    title          = "{LVK} {A}lgorithm {L}ibrary - {LALS}uite",
    howpublished   = "Free software (GPL)",
    doi            = "10.7935/GT1W-FZ16",
    year           = "2018",
	url = {https://git.ligo.org/lscsoft/lalsuite}
}

@article{swiglal_2020,
    title     = "{SWIGLAL}: {P}ython and {O}ctave interfaces to the
                 {LALSuite} gravitational-wave data analysis libraries",
    author    = "Karl Wette",
    journal   = "SoftwareX",
    volume    = "12",
    pages     = "100634",
    year      = "2020",
    doi       = "10.1016/j.softx.2020.100634"
}

@misc{pyfstat,
  author       = {Ashton, Gregory and
                  Keitel, David and
                  Prix, Reinhard and
                  Tenorio, Rodrigo and
                  Ferrer, Maria-Antonia},
  title        = {{PyFstat}},
  month        = 01,
  year         = 2025,
  publisher    = {Zenodo},
  doi          = {10.5281/zenodo.3967045},
  url          = {https://doi.org/10.5281/zenodo.3967045},
  note         = {\url{https://doi.org/10.5281/zenodo.3967045}}
}

@article{jaranowski_data_1998,
	title = {Data analysis of gravitational-wave signals from spinning neutron stars: {The} signal and its detection},
	volume = {58},
	copyright = {http://link.aps.org/licenses/aps-default-license},
	issn = {0556-2821, 1089-4918},
	shorttitle = {Data analysis of gravitational-wave signals from spinning neutron stars},
	url = {https://link.aps.org/doi/10.1103/PhysRevD.58.063001},
	doi = {10.1103/PhysRevD.58.063001},
	number = {6},
	urldate = {2026-03-04},
	journal = {Physical Review D},
	author = {Jaranowski, Piotr and Królak, Andrzej and Schutz, Bernard F.},
	month = {08},
	year = {1998},
	pages = {063001}
}

@article{drago_coherent_2021,
	title = {coherent {WaveBurst}, a pipeline for unmodeled gravitational-wave data analysis},
	volume = {14},
	issn = {23527110},
	url = {https://linkinghub.elsevier.com/retrieve/pii/S2352711021000236},
	doi = {10.1016/j.softx.2021.100678},
	urldate = {2026-03-04},
	journal = {SoftwareX},
	author = {Drago, Marco and Klimenko, Sergey and Lazzaro, Claudia and Milotti, Edoardo and Mitselmakher, Guenakh and Necula, Valentin and O’Brian, Brendan and Prodi, Giovanni Andrea and Salemi, Francesco and Szczepanczyk, Marek and Tiwari, Shubhanshu and Tiwari, Vaibhav and V, Gayathri and Vedovato, Gabriele and Yakushin, Igor},
	month = {06},
	year = {2021},
	pages = {100678},
}

@article{sutton_x-pipeline_2010,
	title = {X-{Pipeline}: an analysis package for autonomous gravitational-wave burst searches},
	volume = {12},
	issn = {1367-2630},
	shorttitle = {X-{Pipeline}},
	url = {https://iopscience.iop.org/article/10.1088/1367-2630/12/5/053034},
	doi = {10.1088/1367-2630/12/5/053034},
	number = {5},
	urldate = {2026-03-04},
	journal = {New Journal of Physics},
	author = {Sutton, Patrick J and Jones, Gareth and Chatterji, Shourov and Kalmus, Peter and Leonor, Isabel and Poprocki, Stephen and Rollins, Jameson and Searle, Antony and Stein, Leo and Tinto, Massimo and Was, Michal},
	month = {05},
	year = {2010},
	pages = {053034}
}

@article{was_performance_2012,
	title = {Performance of an externally triggered gravitational-wave burst search},
	volume = {86},
	copyright = {http://link.aps.org/licenses/aps-default-license},
	issn = {1550-7998, 1550-2368},
	url = {https://link.aps.org/doi/10.1103/PhysRevD.86.022003},
	doi = {10.1103/PhysRevD.86.022003},
	number = {2},
	urldate = {2026-03-04},
	journal = {Physical Review D},
	author = {Wąs, Michał and Sutton, Patrick J. and Jones, Gareth and Leonor, Isabel},
	month = {07},
	year = {2012},
	pages = {022003}
}

@misc{sutton_rule_2013,
	title = {A {Rule} of {Thumb} for the {Detectability} of {Gravitational}-{Wave} {Bursts}},
	copyright = {arXiv.org perpetual, non-exclusive license},
	url = {https://arxiv.org/abs/1304.0210},
	doi = {10.48550/ARXIV.1304.0210},
	urldate = {2026-03-04},
	publisher = {arXiv},
	author = {Sutton, Patrick J.},
	year = {2013},
	note = {Version Number: 1},
}

@article{srivastava_science-driven_2022,
	title = {Science-driven {Tunable} {Design} of {Cosmic} {Explorer} {Detectors}},
	volume = {931},
	issn = {0004-637X, 1538-4357},
	url = {https://iopscience.iop.org/article/10.3847/1538-4357/ac5f04},
	doi = {10.3847/1538-4357/ac5f04},
	number = {1},
	urldate = {2026-03-04},
	journal = {The Astrophysical Journal},
	author = {Srivastava, Varun and Davis, Derek and Kuns, Kevin and Landry, Philippe and Ballmer, Stefan and Evans, Matthew and Hall, Evan D. and Read, Jocelyn and Sathyaprakash, B. S.},
	month = {05},
	year = {2022},
	pages = {22}
}

@article{capote_advanced_2025,
	title = {Advanced {LIGO} detector performance in the fourth observing run},
	volume = {111},
	issn = {2470-0010, 2470-0029},
	url = {https://link.aps.org/doi/10.1103/PhysRevD.111.062002},
	doi = {10.1103/PhysRevD.111.062002},
	number = {6},
	urldate = {2026-03-04},
	journal = {Physical Review D},
	author = {Capote, E. and others},
	month = {03},
	year = {2025},
	pages = {062002}
}

@misc{evans_horizon_2021,
	title = {A {Horizon} {Study} for {Cosmic} {Explorer}: {Science}, {Observatories}, and {Community}},
	copyright = {arXiv.org perpetual, non-exclusive license},
	shorttitle = {A {Horizon} {Study} for {Cosmic} {Explorer}},
	url = {https://arxiv.org/abs/2109.09882},
	doi = {10.48550/ARXIV.2109.09882},
	urldate = {2026-03-04},
	publisher = {arXiv},
	author = {Evans, Matthew and others},
	year = {2021},
	note = {Version Number: 2}
}

@misc{Aplus_GWINC_2018,
  title  = {{The A+ design curve}},
  howpublished = {LIGO Document Control Center},
  author = {L Barsotti and P Fritschel and M Evans and S Gras},
  year   = {2018},
  url    = {https://dcc.ligo.org/LIGO-T1800042/public},
  number = {LIGO-T1800042},
  institution = {LIGO Laboratory}
}

@misc{Aplus_GWINC,
  title  = {{A+/O5 strain curve projections}},
  howpublished = {LIGO Document Control Center},
  author = {Fritschel, Peter and Kabagoz, Begum and Kuns, Kevin and Effler, Anamaria},
  year   = {2025},
  url    = {https://dcc.ligo.org/LIGO-T2500310/public},
  number = {LIGO-T2500310},
  institution = {LIGO Laboratory}
}

@article{andersson_towards_1998,
	title = {Towards gravitational wave asteroseismology},
	volume = {299},
	issn = {0035-8711, 1365-2966},
	url = {https://academic.oup.com/mnras/article/299/4/1059/1108897},
	doi = {10.1046/j.1365-8711.1998.01840.x},
	number = {4},
	urldate = {2026-03-16},
	journal = {Monthly Notices of the Royal Astronomical Society},
	author = {Andersson, N. and Kokkotas, K. D.},
	month = {10},
	year = {1998},
	pages = {1059--1068},
}

@article{buchdahl_general_1959,
	title = {General {Relativistic} {Fluid} {Spheres}},
	volume = {116},
	copyright = {http://link.aps.org/licenses/aps-default-license},
	issn = {0031-899X},
	url = {https://link.aps.org/doi/10.1103/PhysRev.116.1027},
	doi = {10.1103/PhysRev.116.1027},
	number = {4},
	urldate = {2026-03-16},
	journal = {Physical Review},
	author = {Buchdahl, H. A.},
	month = {11},
	year = {1959},
	pages = {1027--1034},
}

@article{regimbau_mock_2026,
	title = {Mock data challenge for next-generation detectors},
	volume = {113},
	issn = {2470-0010, 2470-0029},
	url = {https://link.aps.org/doi/10.1103/d66c-35dy},
	doi = {10.1103/d66c-35dy},
	number = {6},
	urldate = {2026-03-27},
	journal = {Physical Review D},
	author = {Regimbau, Tania and Suresh, Jishnu},
	month = {03},
	year = {2026},
	pages = {063028},
}

@article{regimbau_mock_2012,
	title = {Mock data challenge for the {Einstein} {Gravitational}-{Wave} {Telescope}},
	volume = {86},
	copyright = {http://link.aps.org/licenses/aps-default-license},
	issn = {1550-7998, 1550-2368},
	url = {https://link.aps.org/doi/10.1103/PhysRevD.86.122001},
	doi = {10.1103/PhysRevD.86.122001},
	number = {12},
	urldate = {2026-03-27},
	journal = {Physical Review D},
	author = {Regimbau, Tania and Dent, Thomas and Del Pozzo, Walter and Giampanis, Stefanos and Li, Tjonnie G. F. and Robinson, Craig and Van Den Broeck, Chris and Meacher, Duncan and Rodriguez, Carl and Sathyaprakash, B. S. and Wójcik, Katarzyna},
	month = {12},
	year = {2012},
	pages = {122001},
}

@misc{meacher_einstein_telescope_2014,
  author = {Meacher, Duncan},
  howpublished = {LIGO Document Control Center},
  title = {Einstein Telescope pre site selection detector locations},
  year = {2014},
  url = {https://dcc.ligo.org/LIGO-T1400308},
  number = {LIGO-T1400308},
  institution = {LIGO Laboratory}
}

@misc{ET_coba,
  author = {Danilishin, Stefan and Zhang, Teng},
  title = {{ET} sensitivity curves used for {CoBA} Science Study},
  year = {2023},
  url = {https://apps.et-gw.eu/tds/?r=18213}
}

@article{branchesi_science_2023,
	title = {Science with the {Einstein} {Telescope}: a comparison of different designs},
	volume = {2023},
	issn = {1475-7516},
	shorttitle = {Science with the {Einstein} {Telescope}},
	url = {https://iopscience.iop.org/article/10.1088/1475-7516/2023/07/068},
	doi = {10.1088/1475-7516/2023/07/068},
	number = {07},
	urldate = {2026-03-27},
	journal = {Journal of Cosmology and Astroparticle Physics},
	author = {Branchesi, Marica and others},
	month = {07},
	year = {2023},
	pages = {068},
}

@misc{bruno_einstein_2026,
	title = {Einstein {Telescope}: the triangle in a network and the impact of the relative orientation of detectors},
	copyright = {Creative Commons Attribution 4.0 International},
	shorttitle = {Einstein {Telescope}},
	url = {https://arxiv.org/abs/2610.17446},
	doi = {10.48550/ARXIV.2610.17446},
	urldate = {2026-10-15},
	publisher = {arXiv},
	author = {Bruno, Giacomo and Oswald, Antonin and De Lillo, Federico},
	month = 10,
	year = {2026},
	note = {Version Number: 1},
}

@misc{keane_square_2025,
	title = {A {Square} {Kilometre} {Array} {Pulsar} {Census}},
	copyright = {Creative Commons Attribution 4.0 International},
	url = {https://arxiv.org/abs/2512.16153},
	doi = {10.48550/ARXIV.2512.16153},
	urldate = {2026-04-13},
	publisher = {arXiv},
	author = {Keane, E. F. and Graber, V. and Levin, L. and Tan, C. M. and Johnson, O. A. and Ng, C. and Pardo-Araujo, C. and Ronchi, M. and Vohl, D. and Xue, M. and Group, The SKA Pulsar Science Working},
	year = {2025},
	note = {Version Number: 1},
}

\end{document}